\documentclass[11pt]{article}
\usepackage{epsfig} 
\begin{document}

% ======================= TITLE PAGE ==========================
\begin{titlepage}
\begin{center}

Draft: \today	
{}~             
{}~		

\vskip1.0in 
{\Large\bf Real Time Dynamics of Soliton Diffusion}

\vskip.3in

S. Alamoudi$^{(a)}$\footnote{Email: {\tt smast15@vms.cis.pitt.edu}},
D. Boyanovsky$^{(a)}$\footnote{Email: {\tt boyan@vms.cis.pitt.edu}} and 
F.I. Takakura$^{(a,b)}$\footnote{Email: {\tt takakura@fisica.ufjf.br}}\\ 

\bigskip

{\it (a) Dept.\ of Physics and Astronomy, University of Pittsburgh, 
         Pittsburgh PA USA 15260}\\
{\it (b) Departamento de F\'{\i}sica, Instituto de Ci\^encias Exatas, \\
Universidade Federal de Juiz de Fora, Juiz de Fora, 36036-330, MG, Brasil}\\

\vskip.3in

\end{center}

\vskip.5in

\begin{abstract}
We study the non-equilibrium dynamics of solitons in model Hamiltonians for
Peierls dimerized quasi-one dimensional conducting polymers and commensurate
charge density wave systems. The {\em real time} equation of motion for the
collective coordinate of the soliton and the associated Langevin equation is 
found in a consistent adiabatic expansion in terms of the ratio of the optical 
phonon  or phason frequency to the soliton mass. The equation of motion for the
soliton collective coordinate allows to obtain the frequency dependent soliton
conductivity. 
In lowest order we find that although the coefficient of {\em static} friction
vanishes, there is dynamical dissipation represented by a non-Markovian
dissipative kernel associated with two-phonon processes. The correlation
function of the noise in the quantum Langevin equation and the dissipative
kernel are related by a generalized quantum fluctuation dissipation relation.
To lowest adiabatic order we find that the noise is gaussian, additive and
colored.
We numerically solve the equations of motion in lowest adiabatic order and 
compare to the Markovian approximation which is shown to fail both in the 
$\phi^4$ and the Sine Gordon models even at large temperatures. 

\end{abstract}

\end{titlepage}

\setcounter{footnote}{0}

% ====================  TEXT BEGINS ===========================

\section{Introduction and Motivation}
Since the original work of Krumhansl and Schrieffer\cite{krum} on solitons
as excitations in quasi-one dimensional systems, it has been realized that
solitons play a fundamental role in the transport properties of quasi-one
dimensional Peierls dimerized conducting polymers\cite{su,schriefrev,yu} and
commensurate charge density wave systems\cite{d3}-\cite{horo}.

An important line of experimental and theoretical study has been to determine
the dissipative aspects of soliton dynamics\cite{schriefrev,rev}. 
Soliton diffusion may play an important role in the dynamics of 
photoexcitations, in the photoconductivity in conducting polymers and in the 
transport phenomena associated with phase solitons in charge density wave 
systems\cite{d3}-\cite{horo}.

Early numerical simulations of classical model Hamiltonians 
revealed\cite{several} that solitons undergo Brownian-like motion. A study of
the interaction of solitons with phonon wave-packets showed that
wave-packet-soliton collisions  induce a random-like motion of the
soliton\cite{wada}. One of the main focus of study was the determination of
the diffusion constant which was estimated in\cite{wada} for the $\phi^4$ model
Hamiltonian and in\cite{maki} for the continuum model of 
trans-polyacetylene\cite{tlm}. In these studies the process of soliton 
scattering  off optical\cite{wada,maki} and acoustical\cite{maki} phonons  was
studied and input in (semi) classical
estimates of the diffusion constant based on the classical fluctuation 
dissipation theorem.  A more microscopic formulation of the
calculation of friction and diffusion coefficients of solitons based on
the linear response analysis in terms of Mori's formulation was presented 
in\cite{ogata}. These authors focused on obtaining the {\em static} friction
coefficient by evaluating the correlation functions of the soliton velocity and
using Mori's formulation. 

 There are very few experimental data available on 
the dynamics of soliton diffusion. Although neutral soliton diffusion has been
observed (for a thorough review see\cite{schriefrev,rev}) the main dependence
seems to be determined by soliton trapping and pinning. Thus the experimental
evidence for soliton diffusion is at best inconclusive.

Recently a more microscopic approach to the study of the non-equilibrium
aspects of soliton dynamics has been proposed\cite{neto}.  This approach is
based on the treatment of particle-reservoir models in which the soliton is
taken to be the particle and the phonon fluctuations as the reservoir. The
phonon degrees of freedom are ``integrated out'' in a perturbative manner 
leading to a non-equilibrium effective action of the soliton. 

In this article we study the non-equilibrium dynamics of the soliton following 
this latest approach applied to microscopic models relevant to the description 
of solitons.
The soliton dynamics is treated via the collective
coordinate method in which the coordinate representing the center of mass of 
the soliton becomes a quantum mechanical variable. 

The novel aspect of this article as compared to previous work is that
we use the Schwinger-Keldysh\cite{b1}-\cite{b7}
formulation of non-equilibrium statistical mechanics to obtain the 
{\em real time} equations
of motion for this collective coordinate and the corresponding Langevin
equation by tracing out the phonon degrees of freedom in a consistent adiabatic
expansion in the ratio of the optical phonon (phason) frequency to the soliton 
mass. 

This Langevin equation allows the unambiguous identification of the dissipative
kernels and the noise correlation function thus allowing to establish a 
generalized fluctuation dissipation relation.

 To lowest order in the adiabatic
expansion we find that the dissipative kernels have memory and a Markovian
approximation is unreliable. For the materials of interest, such as 
polyacetylene  with
typical optical phonon frequency $\approx 0.1-0.2 \mbox{ eV }$ or charge 
density wave systems with a typical phason frequency 
$\approx 10^{-3} \mbox{ eV }$ and band-widths of several eV, we find that the 
classical limit of the generalized fluctuation dissipation theorem is not 
applicable. 

To our knowledge none of the previous approaches to soliton dynamics focused on
obtaining the real-time equations of motion, and its solutions in particularly 
relevant cases, nor on the quantum Langevin equation and the properties of the 
stochastic noise and the quantum fluctuation dissipation relation.  

In section 2 we introduce and motivate the models to be studied and determine
the  range of parameters that are experimentally relevant. Section 3 summarizes
the relevant aspects of collective coordinate quantization as applied to
the problems under study. Section 4 presents the non-equilibrium formulation 
for obtaining the equations of motion and the Langevin equation in the general 
case and discusses the features of the solution and the generalized quantum 
fluctuation dissipation relation between the noise correlation function and the
dissipative kernel. Section 5 studies specific model
Hamiltonians: the $\phi^4$ and Sine-Gordon field theories, and analyzes the 
Markovian approximation and the validity of the classical limit. Section 6 
summarizes our conclusions, poses further questions and possible future 
directions.

\section{The models:}

Although we are primarily interested in studying non-equilibrium soliton dynamics 
in quasi-one dimensional conducting polymers such as {\em trans}-polyacetylene 
and charge density wave systems which are electron-phonon systems, we will
use microscopic model Hamiltonians that are somewhat simpler to study. In what
follows we will set for convenience $\hbar = k_B =1$.
For polyacetylene in principle we should start our analysis from the continuum
model of Takayama, Lin-Liu and Maki\cite{tlm}, however as these authors showed
the soliton in this continuum model are similar to those of the Hamiltonian model
studied by Krumhansl and Schrieffer\cite{krum}. 
In particular Ogata et.al.\cite{ogata} had previously used the $\phi^4$ field
theory as model Hamiltonians for conducting polymers. 
Thus we will study the
simpler microscopic model defined by the Hamiltonian\cite{krum}

\begin{equation}
H= \int \frac{dx}{l}\left[ \frac{P^2(x)}{2m} + \frac{A}{2} u^2(x) + \frac{B}{4} 
u^4(x) + \frac{mc^2_o}{2} \left(\frac{du(x)}{dx}\right)^2 \right].
\label{modelhamfi4} 
\end{equation}

Clearly the quantitative details of the dissipative processes in this model 
will be different from those of the continuum model since this model does not
incorporate electrons. However we expect the qualitative features to be
robust. Upon rescaling length and time scales,  performing a canonical 
transformation and  adding suitable constants, the
Hamiltonian (\ref{modelhamfi4}) obtains the form of a $\phi^4$ field theory 
\begin{equation}
H = \int dx \left[\frac{\Pi^2}{2} + \frac{1}{2} \left(\frac{d \phi}{dx}\right)^2+
U(\phi) \right] \; \; ; \; \; U(\phi) = \frac{1}{2g}(m^2-g\phi^2)^2,
\label{fi4}
\end{equation}
\noindent where the constants $m ; g$ are determined by the original parameters in
(\ref{modelhamfi4}). These two parameters can be related to the optical phonon
frequency $\omega_o = 2m $ and the soliton mass $M=\frac{4m^3}{3g}$ 
(see next section).
As it will become clear in the following sections, the equation of motion can
be obtained in a systematic expansion in the adiabatic ratio $m/M$ which
is identified with the dimensionless coupling constant $g/m^2$ of the 
field theory. In this
model we identify the soliton mass with the rest energy of the soliton, and
using the parameters for {\em trans}-polyacetylene given by\cite{su,schriefrev,yu}
$\omega_o \approx 0.12 \mbox{ eV }\; ; \; M \approx 0.4 \mbox{ eV }$  we find 
that  the adiabatic ratio $m/M \approx 0.15$ is  small and a perturbative 
expansion in this ratio may be appropriate. 

In charge density wave systems, beginning from the Landau-Ginzburg description 
of the quasi-one dimensional system\cite{d3}-\cite{horo}, and fixing the 
amplitude of the order parameter (gap) but allowing the phase to fluctuate, the
dynamics is determined by the effective Hamiltonian for the phase of the order 
parameter (for details see\cite{d3}-\cite{horo})
\begin{equation}
H= \frac{n(\epsilon_F)}{4}\int dx \left[v^2_f \left( \frac{d\phi}{dx}\right)^2+
\frac{m^*_e}{m_e}\left(\frac{d\phi}{dt}\right)^2+\frac{\omega^2_F m^*_e}{
{\cal M}^2 m_e}\cos({\cal M}\phi)\right] \label{cdwham}
\end{equation}
with $m_e \; ; \; m^*_e$ the electron's mass and its effective mass, $\omega_F 
\; ; \; v_F$ the Fermi frequency and velocity and ${\cal M}$ the commensurability
of the charge density wave\cite{d3}-\cite{horo}. Again after suitable rescalings
of time and space and a canonical transformation, the Hamiltonian can be cast
as in eqn.(\ref{fi4}) above but with the potential given by
\begin{equation}
U(\phi) = U(\phi) = {m^4\over g} \left( 1 - \cos\left[ {\sqrt g \over m} 
\phi \right] \right).
\label{singor}
\end{equation}
In this model the gap in the phason spectrum is identified with $m$ and
the soliton energy is given by $M= 8 m^3/g$. For a typical material, such as
$K_{0.3}M_o0_3$ the gap in the phason spectrum is $\approx 10^{-3}
 \mbox{ eV }$
whereas the soliton energy is 
$\approx 3 \times 10^{-2} \mbox{ eV }$\cite{d3}-\cite{horo}. Therefore
for this type of materials the adiabatic ratio $m/M \approx 0.03$ and a 
perturbative expansion is reliable. 

We must note that for both cases the temperatures of experimental relevance
correspond to $T < 3 \times 10^{-2} \mbox { eV }$ which are of the order of 
(or smaller than) the typical optical phonon or phason frequencies and the
nature of a classical limit must be understood carefully.

\section{Collective Coordinate Quantization:}
In the previous section we have provided a rational for studying the
dynamics of solitons in model field theories described by Hamiltonians of
the form
\begin{equation}
H\,=\,\int dx\,\left\{\frac{\pi^2}{2}+\frac{1}{2}\left(\frac{d\phi}{dx}\right)^2
+U(\phi)\right\}
\label{hs}
\end{equation}
after suitable rescaling of the parameters. 

A static soliton is a solution of the time independent field equation 
\begin{equation}
-\frac{d^2\phi_s}{dx^2}+\frac{\partial U(\phi_s)}{\partial\phi}\,=\,0 \label{eqnm}
\end{equation}
with boundary conditions such that $\vert x \vert \rightarrow \infty \, , \, 
\phi_s \rightarrow \pm 
\phi_\infty \hbox{ and } U(\phi_\infty) = 0$ \cite{c7,d2}.
Translational invariance implies that such solution is of the form
$\phi_s(x-x_0)$ with $x_0$ an arbitrary translation chosen such that 
$\phi_s(0)=0$, therefore $x_0$ is identified with the position of the soliton.

Including the time derivatives in the equations of motion one sees that
after (proper rescalings of time and space that led to the form of the
Hamiltonian given above) they are invariant under a ``Lorentz'' transformation. 
A soliton moving with constant velocity is given by $\phi_s\left[
\frac{x-x_0-vt}{\sqrt{1-v^2}}\right]$\cite{krum,wada,d3}. The energy of a static
soliton is
identified in these models with the soliton mass, $M$, and is given by
\begin{equation}
M \equiv E[\phi_s] = \int {\rm d} x \left({{{\rm d} \phi_s}\over{{\rm d} x}}
\right)^2.
\label{M}
\end{equation}

Quantization around the static soliton solution implies writing
\begin{equation}
\hat{\phi}(x,t)\,=\,\phi_s(x - x_0)+\hat{\psi}(x - x_0;t). \label{ex}
\end{equation}
Where the fluctuation operator is expanded in terms of a complete set of 
harmonic modes around the soliton 

\begin{equation}
\hat{\psi}(x-x_0;t)\,=\,\sum_{n}^\infty q_n(t)\, {\cal U}_n(x - x_0)
\label{fluk}
\end{equation}

\noindent where the mode functions  ${\cal U}_n(x - x_0)$  obey

\begin{equation}
\left[- {{\rm d}^2\over{{\rm d} x^2}} + \left.\frac{d^2U}{d\phi^2}\right|_{\phi_s}
\right] 
{\cal U}_n(x - x_0) = \omega_n^2 \, {\cal U}_n(x - x_0) 
\label{hosc}
\end{equation}

 \noindent with the completeness relation given by
\begin{equation}
\sum_{\mbox{b}} \;{\cal U}^*_b(x - x_0) \, {\cal U}_b(x^\prime - x_0) + 
\int dk \;{\cal U}^*_k(x - x_0) \, {\cal U}_k(x^\prime - x_0)\,=\,
\delta(x - x^\prime)
\end{equation}
and the subscript b stands for summation over bound states and k for
scattering states. For bound states, the eigenvectors are chosen to be real and
for scattering states, we label them as ${\cal U}_k(x - x_0)$ and
are chosen such that ${\cal U}_k^\ast = {\cal U}_{-k}$, in which case the 
coordinate operators obey the hermiticity condition $q^*_k(t) = q_{-k}(t)$.

These eigenvectors are normalized as
\begin{equation}
\int {\rm d} x \, {\cal U}^\ast_p(x - \hat x_0)\, {\cal U}_q(x - \hat x_0) =
\delta_{p,q}. 
\label{upuq}
\end{equation}

As a consequence of translational invariance, there is a mode with zero
eigenvalue given by\cite{c7}

\begin{equation} 
{\cal U}_0(x - x_0) = {1 \over{\sqrt M}} \left({{{\rm d} \phi_s}\over{{\rm d} x}}
\right) .
\label{zero}
\end{equation}

Depending on the particular form of the potential $U(\phi)$ there may
be other bound states (as is the case with the $\phi^4$ potential). There
is a continuum of scattering states with frequencies $\omega_k^2= k^2+\omega^2_o
\; ; \; \omega^2_o = d^2U(\phi)/d^2\phi |_{\phi_{\infty}}$. These scattering 
states correspond  asymptotically to phase shifted plane waves in the cases 
under consideration because the relevant potentials are reflectionless\cite{c7,d2}.
The frequencies
$\omega_o$ are identified with the optical phonon frequencies in the case of
the $\phi^4$ model\cite{krum,wada,ogata} and of the phason gap in the case of 
phase solitons in charge density wave systems\cite{d3}-\cite{horo}.

The fluctuation along the functional direction corresponding to the zero 
frequency mode  represents an infinitesimal translation of the soliton that
costs no energy. Since this
mode has no restoring force, any arbitrarily large amplitude fluctuation 
along this direction is energetically allowed. Therefore fluctuations along
this direction must be treated non-perturbatively. The variable $x_0$, i.e. the
center of mass of the soliton is elevated to the status of a quantum mechanical
variable, and the fluctuations are orthogonal to the zero mode. 
This treatment is the basis of the collective coordinate 
method\cite{c1}-\cite{c22} which was previously used within the context of 
soliton dynamics by
Wada and Schrieffer\cite{wada}, Maki\cite{maki}, Ogata et. al.\cite{ogata} 
and within the context of polaron dynamics by Holstein and Turkevich\cite{c6}. 
More recently Caldeira and Castro Neto implemented the collective coordinate
quantization method combined with influence functional techniques for the
treatment of solitons and polarons\cite{neto}.

In collective coordinates quantization  instead of the expansion (\ref{ex}) with
(\ref{fluk}) we expand ${\phi}(x,t)$ as 
\begin{equation}
\phi(x,t) = \phi_s(x - \hat x_0(t)) + \sum_{n \neq 0}^\infty Q_n(t)\, 
{\cal U}_n(x - \hat x_0(t) )\;.
\label{phicc}
\end{equation}

This amounts to a change of basis in functional space, from the ``cartesian''
coordinates $\{ q_n \}$ to ``curvilinear'' coordinates $\{\hat{x}_0,Q_{n\neq 0} \}$. 

The next step is to express the Hamiltonian in terms of the new variables 
$\hat{x}_0(t)$ and $Q_n(t)$. For this we find more clear and convenient
the analysis presented by Holstein and Turkevich\cite{c6} which we summarize
below for the cases under consideration. 

\subsection{ The Kinetic and Potential Energies }

In the Schroedinger representation the kinetic energy can be expressed as
a functional derivative as
\begin{equation}
T = -{1\over 2} \int {\rm d} x {\delta\over{\delta \phi}} {\delta 
\over{\delta \phi}}\; ,
\label{Tphi}
\end{equation}
where the functional derivative is written in the new coordinates using the 
chain-rule
\begin{equation}
{\delta \over{\delta \phi(x)}} = {\delta \hat x_0 \over{\delta \phi(x)}} {\delta 
\over{\delta \hat x_0}} + \sum_{m \neq 0} {\delta Q_m \over{\delta \phi(x)}} 
{\delta \over{\delta Q_m}}.
\label{ddelphi}
\end{equation}

Taking the functional variation of the field $\phi$, eq.(\ref{phicc}), we obtain
\begin{eqnarray}
\delta \phi(x) &=& {\delta \phi(x) \over{\delta \hat x_0}} \delta \hat x_0 +
\sum_{m \neq 0} {\delta \phi(x) \over{\delta Q_m}} \delta Q_m \nonumber \\
 &=& \left[ {\partial \phi_s(x - \hat x_0) \over{\partial 
\hat x_0}} +
\sum_{m\neq 0} Q_m {\partial {\cal U}_m(x - \hat x_0) \over{\partial \hat x_0}}
\right] 
\delta \hat x_0 + \sum_{n \neq 0} {\cal U}_n(x - \hat x_0) \delta Q_n \; . 
\label{delphi}
\end{eqnarray}

Projecting both sides of the above equation on  ${\cal U}^\ast_0(x - \hat x_0)$ 
and then  ${\cal U}^\ast_p(x - \hat x_0)$ with $p\not=0$, using eqn.(\ref{zero})
and the orthonormalization condition eqn.(\ref{upuq}), we obtain: 
\begin{eqnarray}
{\delta \hat x_0 \over{\delta \phi(x)}} & =&  - {1\over{\sqrt M}} 
{1\over{\left[ 1 + (1/\sqrt M) \sum_{m \neq 0} Q_m S_m \right]}} 
{\cal U}^\ast_0(x - \hat x_0) \label{xx} \\
{\delta Q_p \over{\delta \phi(x)}}& = &  {\cal U}^\ast_p(x - \hat x_0) - 
{1\over{\sqrt M}} {{\sum_{n \neq 0} G_{pn}Q_n}\over{\left[ 1 
+ (1/\sqrt M) \sum_{m \neq 0} Q_m S_m \right]}} 
{\cal U}^\ast_0(x - \hat x_0), \label{cddphi}
\end{eqnarray}
where the matrix elements $G_{pm}$ are defined as
\begin{equation}
G_{pm} \,=\, \int {\rm d} x \,{\cal U}^\ast_p(x - \hat x_0)\,
\frac{\partial{\cal U}_m(x - \hat x_0)}{\partial x} \label{me1} 
\end{equation}
\begin{equation}
S_m \,\equiv\, G_{0m} =  \int {\rm d} x \,{\cal U}_0(x - \hat x_0) \,
\frac{\partial{\cal U}_m(x - \hat x_0)}{\partial x} \; .
\label{gmn}
\end{equation}

At this stage it is straightforward to follow the procedure detailed
in\cite{c6} to find the final form of the kinetic term in the Hamiltonian
in the Schroedinger representation of the coordinates 
$\hat{x}_0, Q_{m \neq 0}$:

\begin{eqnarray}
T &=& - {1\over 2} \left\{ {1\over{D}} {\delta \over{\delta \hat x_0}}
{\delta \over{\delta \hat x_0}} + {1\over{\sqrt D}} 
{\delta \over{\delta \hat x_0}} \sum_{p,m \neq 0}\left[ 
\frac{G_{pm}Q_m}{\sqrt{D}} {\delta \over{\delta Q_p}} + {\delta \over{\delta Q_p}}
\frac{G_{pm}Q_m}{\sqrt{D}} \right] + \right. \nonumber \\
&& \quad\quad \left. {1\over{\sqrt D}} \sum_{p,q,m,n\neq 0} 
{\delta \over{\delta Q_p}}\left[\delta_{-p,q} \sqrt{D}\,+\,
\frac{G_{pm}Q_m}{\sqrt{D}}G_{qn}Q_n
\right]{\delta \over{\delta Q_q}} \right\},
\label{Tqxf}
\end{eqnarray}

\noindent where $\sqrt D$ is the Jacobian associated with the change of 
coordinates\cite{c6,c7} and given by
\begin{equation}
\sqrt D \equiv \sqrt M \left[ 1+ \frac{1}{\sqrt M}\sum_{m \neq 0}Q_m S_m \right]. 
\label{jacobian}
\end{equation}

The total potential energy, including the elastic term, $V[\phi ]$ (see 
eqn.(\ref{hs})), is given by
\begin{equation}
V[\phi ] \equiv \int {\rm d} x \left[ {1\over 2} 
  \left({{\partial \phi}\over{\partial x}}\right)^2 + U(\phi) \right]\, .
\end{equation}

Using the expansion given by eqn.(\ref{phicc}) we find that it can be written 
in terms of the new coordinates as
\begin{equation}
V[\phi] = M + {1\over 2} \sum_{m \neq 0} Q_m Q_{-m} \omega_m^2 + {\cal O}(Q^3) 
+\cdots \,.\label{Vfin}
\end{equation}
By translational invariance the potential energy does not depend on the
collective coordinate. 
Identifying the canonical momenta conjugate to $\hat{x}_0, Q_n$ as 
\begin{equation}
\pi_0 \equiv P\;=\;-i {\delta \over{\delta \hat x_0}} \quad ; \quad
\pi_k = -i {\delta \over{\delta Q_{-k}}} \quad \hbox{for} \quad k \neq 0 \,.
\label{ppi}
\end{equation}

\noindent and using  the commutation relation of $\sqrt D \hbox{ and } 1/\sqrt D 
\,$ with $Q_n, \, \pi_n \hbox{ and } P$ given by
\begin{equation}
\left[\pi_n, \sqrt D\right] = -i S_n \quad \hbox{and} \quad ; \quad \left[\pi_n, 
{1\over{\sqrt D}}\right] = -i {S_n \over D}, 
\label{pisd}
\end{equation}
 we find the final form of the Hamiltonian:

\begin{eqnarray}
H &=& M + {1\over 2} \left\{ {P^2\over{D}} + {P\over{\sqrt D}} \sum_{p,m \neq 0}
\left[ 
\frac{G_{pm}Q_m}{\sqrt{D}} \pi_{-p} + \pi_{-p} \frac{G_{pm}Q_m}{\sqrt{D}} \right]
+ \sum_{p \neq 0} \omega_p^2\,Q_p Q_{-p} \right. + \nonumber \\
&& \quad\quad \left. {1\over{\sqrt D}} \sum_{p,q,m,n\neq 0} 
\pi_{-p}\left[\delta_{-p,q} \sqrt{D}\,+\,\frac{G_{pm}Q_m}{\sqrt{D}}G_{qn}Q_n
\right]\pi_{-q} \right\} + {\cal O}(Q^3)+\cdots,
\end{eqnarray}
where $Q_p$ are now operators. The coordinates $Q_k$ associated with
the scattering states describe optical phonon (or phason)  degrees of freedom 
with the optical phonon (phason) frequency 
$\omega_o = d^2U(\phi)/d^2\phi|_{\phi_{\infty}}$. 
Since the Hamiltonian does not depend on $\hat{x}_0$ its canonical momentum $P$
is conserved, it is identified with the total momentum of the soliton-phonon 
(phason) system\cite{c6,c7}. The soliton velocity, however, is not proportional
to $P$ and depends on the momentum of the phonon (phason) field. 

\subsection{Coupling to external fields:}

The main goal of studying the non-equilibrium dynamics of soliton is a deeper
understanding of transport processes by these topological excitations. In the
case of conducting polymers in which the underlying physics is described by
electron-phonon interactions, the soliton excitation in the dimerized state
induces a fractionally charged state associated with an electronic bound state
in the middle of the electronic gap\cite{jackiw}. The charge density associated
with the electronic bound state is proportional to the profile of the phonon
zero mode given by (\ref{zero}) i.e. $\rho_c(x,t) = e{\cal C} {\cal U}_0(x-x_0(t))$
which is localized at the center of mass of the soliton, the constant ${\cal C}$
depends on the (fractional) charge localized around the soliton\cite{jackiw,wil}. 

In the case of charge density waves, the transport current is identified with
the topological current $j_{\mu} \propto \epsilon_{\mu \nu} \partial_{\nu} \phi$
\cite{d3}-\cite{horo}. Therefore in both cases the charge density is associated 
with the translational zero mode. Furthermore, current conservation\cite{wil}
implies that the spatial current is given by ${\cal J}_x(x,t) = e 
{\cal C} \dot{x}_0(t){\cal U}_0(x-x_0(t))$. 
Hence, a spatially constant external electric field
couples to the translational zero mode and introduces a term in the Lagrangian
of the form
\begin{equation}
\delta {\cal L} = - \int dx E(t)x \rho_c(x). \label{exterele}
\end{equation} 
Taking $\rho_c(x)= e {\cal C} {\cal U}_0(x-\hat{x}_0)$ we find that an external
spatially constant electric field induces a linear term in $\hat{x}_0$ in
the Lagrangian as a consequence of the breakdown of translational invariance,
\begin{equation}
\delta {\cal L} = -\tilde{j}(t) \hat{x}_0 \; ; \; \tilde{j} = 
e {\cal C}E(t), \label{linearterm}
\end{equation}
This term is responsible for
accelerating the soliton and changing the total momentum of the system. 

The total (spatially integrated) current transported by the soliton is then given
by 
\begin{equation} 
\int dx {\cal J}_x(x,t) = {\cal C} \Delta \phi \dot{x}_0(t) \label{current}
\end{equation}
with $\Delta \phi = \phi(x=\infty)-\phi(x=-\infty)$. 
The expression given by eqn. (\ref{current}) will allow us to obtain the 
soliton conductivity when the equation of motion for the collective coordinate
is obtained.  

\section{The Soliton in the Phonon Heat Bath}

Our goal is to study the dynamics of a soliton in interaction with the phonons
(or phasons).
This is achieved by obtaining the real-time equations of motion of the collective 
coordinate $\hat{x}_0$ by  treating the phonons (phasons) as a ``bath'' and 
obtaining an influence functional\cite{d5}-\cite{d8} by ``tracing out'' the 
phonon degrees of freedom. 
 We assume that the total density matrix for the
soliton-phonon system decouples at the initial time $t_i$, i.e.
\begin{equation}
\rho(t_i)\,=\,\rho_s(t_i)\,\otimes\,\rho_R(t_i),
\end{equation}
where $\rho_s(t_i)$ is the density matrix of the system which is taken to be 
that of a free particle associated with the collective coordinate of the
soliton, i.e. $\rho_s(t_i)= |x_0><x_0|$ and $\rho_R(t_i)$ is the density matrix 
of the phonon bath and describes harmonic phonons (or phasons) in thermal 
equilibrium at a temperature $T$.

Since the solitons can never be separated from the phonon fluctuations, this 
factorization must be understood to hold in the limit in which the
initial time $t_i \rightarrow -\infty$ with an adiabatic switching-on  the
soliton-phonon interaction. An approximate influence functional has been 
previously obtained
in lowest order in reference\cite{neto}, but our approach is different in the
sense that we obtain the real-time non-equilibrium evolution equations for
the collective coordinate as well as the quantum Langevin equation.
Furthermore, whereas in reference\cite{neto} only processes that conserve the
phonon number were considered, we account for all the two-phonon processes 
consistently to lowest order in the adiabatic expansion.

The time evolution is completely contained in the time dependent density matrix
\begin{equation}
\rho(t)= U(t,t_i)\rho(t_i)U^{-1}(t,t_i)\label{rhotime}
\end{equation}
with $U(t,t_i)$ the time evolution operator. Real time non-equilibrium expectation
values and correlation functions can be obtained via functional derivatives
with respect to sources\cite{b6} of the generating functional\cite{b1}-\cite{b6}
\begin{equation}
Z[j^+,j^-] = Tr U(\infty,-\infty;j^+)\rho_i U^{-1}(\infty,-\infty;j^-)/Tr\rho(t_i),
\label{genefunc}
\end{equation}
where $j^{\pm}$ are sources coupled to the fields. This generating functional  
is readily obtained using the Schwinger-Keldysh method which involves a path 
integral in a complex contour in time\cite{b1}-\cite{b6}: a branch corresponding
to the time evolution forward, a backward branch corresponding to the inverse 
time evolution operator and a branch along the imaginary time axis
from $t_i$ to $t_i-i\beta$ to represent the initial thermal density matrix. 
We will obtain the equation of motion for the soliton collective coordinate
in an expansion of the ``adiabatic'' parameter $m/M \approx \omega_0/M$, as 
discussed in section 2, this
is also the weak coupling limit of the scalar field theories under consideration
\cite{c7}.
As it will be shown explicitly below in the particular cases studied, the 
matrix elements given by eqns.(\ref{me1},\ref{gmn}) will provide the necessary 
powers of $m$. The lowest order in $m/M$ is formally 
obtained by  keeping only the $1/M$ terms in the Hamiltonian and 
neglecting the non-linear ${\cal O}(Q^3)$ terms. Under these approximations,
$1/D \simeq 1/M$ and the Hamiltonian has the following form
\begin{eqnarray}
H =  M \,+ \,{1\over{2 M}} \left( P + \sum_{m,n \neq 0} D_{mn}\pi_m\, Q_n
\right)^2 + \frac{1}{2}\sum_{m \neq 0}\left[ \pi_m \pi_{-m} + \omega_m^2 Q_m 
Q_{-m} \right] + \tilde{j}(t) \hat{x}_0,
\label{Happrof}
\end{eqnarray}
where we define 
\begin{equation}
D_{mn} = G_{-mn}.
\label{Dmn}
\end{equation}

At this point it proves convenient to write the coordinates and momenta 
of the phonons in terms of creation and annihilation operators obeying the
standard Bose commutation relations,
\begin{equation}
Q_k = {1\over{\sqrt{2\omega_k}}} \left[ a_k + a_{-k}^\dagger \right] 
\; \; ; \; \; 
\pi_k = -i \sqrt{{\omega_k \over 2}} \left[ a_k - a_{-k}^\dagger\right]\,.
\label{QP}
\end{equation}
 
The Hamiltonian can be expressed in terms of $a$ and $a^\dagger$ as
\begin{equation}
H = {1\over{2 M}} \left( P + F[a^\dagger, a]\right)^2 + 
\sum_{k \neq 0} \omega_k \left(a_k^\dagger \,a_k + 1/2\right) +
\tilde{j}(t)\hat{x}_0+ M, 
\label{Haa}
\end{equation}
where 
\begin{equation}
F[a^\dagger, a] = \sum_{p,k\not=0} \left[T^{(S)}_{pk}\,\left(a_p\,a_k \,-\,
a^\dagger_{-p}\,
a^\dagger_{-k}\right)\,+\,T^{(A)}_{pk}\,\left(a^\dagger_{-p}\,a_k \,-\,
a^\dagger_{-k}\,
a_{p}\right) \right].
\label{Faa}
\end{equation}

We have made use of the symmetries of the operators and defined the symmetric 
$T^{(S)}_{pk}$ and antisymmetric $T^{(A)}_{pk}$ matrices that
provide the interaction vertices as
\begin{equation}
T_{kp}^{(S)} = {1\over{4 i}} \left[\sqrt{{\omega_k
\over{\omega_p}}} - \sqrt{{\omega_p\over{\omega_k}}}\, 
\right] D_{kp} \quad ; \quad
T_{kp}^{(A)} = {1\over{4 i}} \left[\sqrt{{\omega_k
\over{\omega_p}}} + \sqrt{{\omega_p\over{\omega_k}}}\, 
\right] D_{kp}. \label{scm}
\end{equation}

To use the path integral formulation we need the Lagrangian, which to the 
order that we are working (${\cal O}(m/M)$) and properly accounting for
normal ordering, is given by
\begin{equation}
{\cal L} [\dot{\hat x_0},a,a^\dagger ] = {M\over 2} \dot{\hat x_0}^2 - \dot{\hat
x_0} F[a^\dagger, a] - \sum_{k \neq 0} \omega_k (a_k^\dagger a_k + 1/2) -
\tilde{j}(t)\hat{x}_0- M.
\label{lag}
\end{equation}

The interaction of the collective coordinate and the phonon degrees of freedom
is now clear. Only time derivatives of the collective coordinate couple, a 
consequence of the Goldstone character of the collective coordinate. There
are two processes described by the interaction: i) creation and destruction of
two phonons and ii) scattering of phonons. Whereas the first type can contribute
with the phonons in their ground state, the second can only contribute if 
phonon states are occupied.

Since we have preferred to work in terms of the creation and annihilation
operators it is convenient to write the path integral for the non-equilibrium 
generating functional in the coherent state representation\cite{negele}.
Following the steps presented in\cite{neto,negele}, we find that the
generating functional of non-equilibrium Green's functions is given by

\begin{equation}
{\cal Z}[j^+,j^-] =  \int {\cal D}x^+ \int {\cal D}x^- \int {\cal D}^2 \gamma^+ 
\int {\cal D}^2 \gamma^-  exp\left\{i\int dt \left({\cal L}[\dot{x}^+,
\gamma^{*+},\gamma^+,j^+]\,-\,{\cal L}[\dot{x}^-,\gamma^-,
\gamma^{*-},j^-]\right)\right\}
\label{trgnr}
\end{equation}
with the Lagrangian density defined on each branch given by
\begin{eqnarray} 
{\cal L}[\dot{x}^{\pm},\gamma^{\pm},\gamma^{*\pm},j^{\pm}] & = &  {M \over 2} 
\left( \dot{x}^{\pm}\right)^2  + \sum_{k\neq 0}\left[i\gamma^{*\pm}_k 
\frac{d \gamma^{\pm}_k}{dt} - \omega_k \gamma_k^{*\pm} \gamma_k^{\pm}+\gamma_k^{\pm}
j^{*\pm}_k+ \gamma_k^{*\pm} j^{\pm}_k\right] + x^{\pm} j^{\pm}_o - \nonumber \\
& &  \dot{x}^{\pm} F[\gamma^{* \pm}, \gamma^{\pm}] -\tilde{j}(t)x^{\pm}
\label{lagnoneq}
\end{eqnarray}
and with proper boundary conditions on the fields that reflect the factorized 
initial condition with the phonons in thermal equilibrium.  The signs $\pm$ in 
the above expressions correspond to the fields and sources on
the forward ($+$) and backward $(-)$ branches. The contribution from the branch
along the imaginary time is cancelled by the normalization factor. This is
the non-equilibrium generalization of the coherent state path integrals.
For more details the reader is referred to the literature\cite{neto,negele}.
Non-equilibrium Green's functions are now obtained as functional derivatives
with respect to the sources $j^{\pm}$. There are 4 types of free phonon 
propagators\cite{b1}-\cite{b6}: 
\begin{eqnarray}
< a_k^{\dagger +}(t) a_p^+(t^\prime)> &=& \delta_{k,p} 
e^{-i\omega_k(t^\prime - t)} [\theta(t^\prime - t) + n_k ] \nonumber \\ 
< a_k^+(\tau) a_p^{\dagger +}(\tau^\prime)> &=& \delta_{k,p} 
e^{i\omega_k(t^\prime - t)} [\theta(t - t^\prime) 
+ n_k ] \nonumber \\ 
< a_k^{\dagger (\pm)}(t) a_p^{\dagger (\pm)}(t^\prime)> &=& 0 
\nonumber \\ 
< a_k^{(\pm)}(t) a^{(\pm)}_p(t^\prime)> &=& 0 \\ 
< a_k^{\dagger +}(\tau) a_p^-(\tau^\prime)> &=& \delta_{k,p} 
e^{-i\omega_k(t^\prime - t)} [1 + n_k ] \nonumber \\
< a_k^+(\tau) a_p^{\dagger -}(\tau^\prime)> &=& \delta_{k,p} 
e^{i\omega_k(t^\prime - t)} n_k,  \nonumber 
\label{propaa}
\end{eqnarray}
where $n_k$ is Bose Einstein distribution for phonons of quantum number $k$ and
$<\cdots>$ refer to averages in the initial density matrix. The
$++$ ($--$) propagators correspond to the time-ordered (anti-time-ordered), 
whereas the $\pm \mp$ are linear combinations of the advanced 
and retarded propagators\cite{b6}.

An important point to notice is that 
\begin{equation}
< F[a^\dagger,a]> \,=\,0 \label{faad}
\end{equation}
in the non-interacting case, since it is proportional to $\sum_k D_{k,-k}=0$.

\subsection{The Soliton Equation of Motion}

The equation of motion of the soliton can be derived by expanding $x^{\pm}(t)\,=\,
q(t)\,+\,\xi^{\pm}(t)$ and requiring $<\xi^{\pm}(t)>\,=\,0$ to all orders in  
perturbation theory. 
Imposing the condition $<\xi^+(\tau^\prime)>\,=\,0$, treating the interaction
term  up to second order in perturbation theory  and using eqn.(\ref{faad}), 
we obtain the following equation of motion
\begin{equation}
\int_{-\infty}^{\infty} dt^\prime <\xi^+(t)\dot{\xi}^+(t^\prime)>\left[
 \left\{ M\dot{q}(t^\prime) +
\int_{-\infty}^t dt^{\prime\prime} \Gamma_m(t^\prime-t^{\prime\prime}) 
\dot{q}(t^{\prime\prime}) \right\} + <\xi^+(t)\xi^+(t^\prime)> 
\tilde{j}(t^\prime)\right]\,=\,0 ,
\label{eqnms}
\end{equation}

\noindent where the retarded kernel is given by
\begin{eqnarray}
-i\,\Gamma_m(t-t^{\prime})\theta(t-t^{\prime}) & = & <F[a^{\dagger +}(t),a^+(t)]
\;F[a^{\dagger +}(t^\prime),a^+(t^\prime)]> \nonumber \\
& - & <F[a^{\dagger +}(t),a^+(t)]\;F[a^{\dagger -}(t^\prime).
a^-(t^\prime)]> \label{GA}
\end{eqnarray}

Alternatively this equation of motion may be obtained by computing the
influence functional\cite{d5}-\cite{d8} in second order perturbation theory.
The resulting
influence functional is quadratic in the collective coordinate,  performing
the shift $x^{\pm}(t) = q(t)+\xi^{\pm}(t)$ the above equation of motion is
obtained by requesting that the linear terms in $\xi^{\pm}$ vanish, (there are
two linear terms, both give the same equation of motion).

The kernel $\Gamma_m(t-t^\prime)$ is found by using  eqn.(\ref{GA}) and 
eqn.(\ref{propaa})  and it is given by 
\begin{eqnarray}
\Gamma_m(t - t^\prime) &=& -4\,  \sum_{p,k\not=0} \left\{ T_{pk}^{(S)}\, 
T_{-p-k}^{(S)} (1 \,+ 2 n_p) \,\sin \left[ (\omega_p + \omega_k)
(t - t^\prime)\right] \right. \nonumber \\
&& \left. -  \quad 2 T_{pk}^{(A)} \, T_{-p-k}^{(A)} n_p 
\sin\left[(\omega_p - \omega_k)(t - t^\prime)\right] \right\}.
\label{GTT}
\end{eqnarray}

Performing the integral over $t^\prime$ in eqn.(\ref{eqnms}) by parts, we 
obtain the final form of the equation of motion 
\begin{equation}
M \ddot{q} (t) + \int_{-\infty}^t {\rm d} t^\prime \,
\Sigma_m(t-t^\prime)\dot{q}(t^\prime) = \tilde{j}(t),
\end{equation}
where the non-local kernel is given by
\begin{equation}
\Sigma_m(t-t^\prime) = {\partial \Gamma_m(t-t^\prime) 
\over{\partial t}}= -{\partial \Gamma_m(t-t^\prime) 
\over{\partial t^{\prime}}}\,.
\label{Sigm}
\end{equation} 

Using eqn.(\ref{GTT}) we find the final expression for the kernel $\Sigma_m$:
\begin{eqnarray}
\Sigma_m(t-t^\prime) &=&  -4 \,  \sum_{p,k\not=0} \left\{ T_{pk}^{(S)} \,
T_{-p-k}^{(S)} (1 +  2n_p ) (\omega_p + \omega_k) 
\cos \left[ (\omega_p + \omega_k)(t - t^\prime)\right]  
\right. \nonumber \\
&&\left. - 2\;T_{pk}^{(A)}\,T_{-p-k}^{(A)} n_p \,
(\omega_p - \omega_k) \cos\left[(\omega_p - \omega_k)
(t - t^\prime)\right] \right\}.
\label{sigmf}
\end{eqnarray}
We will see in the next sections that the two kernels $\Sigma_m \; ; \; \Gamma_m$
have very special significance: whereas $\Sigma_m$ is identified with
the real-time retarded self-energy of the collective coordinate, $\Gamma_m$ 
will provide the
coefficient of {\em dynamical friction} in the Markovian approximation.

It is more convenient to express the equation of motion of the soliton in terms
of the velocity
\begin{equation}
M \dot{V} (t) + \int_{-\infty}^t {\rm d} t^\prime \,
\Sigma_m(t-t^\prime)\,V(t^\prime) = \tilde{j}(t)
\label{eqmotm}
\end{equation}
with $\Sigma_m$ given by eqn.(\ref{sigmf}).

The relation (\ref{sigmf}), ensures to this order in the perturbative
expansion,  that
with an adiabatic switching on convergence factor introduced to regularize
the lower limit of the integral and to provide an initial factorization
of the density matrix as $t_i \rightarrow -\infty$ the
total integral of the retarded self-energy kernel vanishes, i.e,
\begin{equation}
\int_{-\infty}^t \Sigma_m(t-t')dt' = 0. \label{zeroint}
\end{equation}

Therefore, in the absence of a driving term that explicitly breaks translational
invariance, i.e. for $\tilde{j}=0$, any  constant velocity of the soliton is a
solution of the equation of motion (\ref{eqmotm}).  This result is a 
consequence of the Galilean invariance of the effective Lagrangian.

\subsection{General Properties of the Solution}

Consider switching on a spatially constant electric field at $t=0$ to
study the linear response of the soliton velocity. Assuming that for $t<0$ the
soliton traveled with a constant velocity $v_0$,
after switching on the electric
field the soliton will accelerate, but it will also transfer energy and
excite the phonon degrees of freedom and this will lead to dissipative
processes. Therefore writing $V(t)=v_0+ v(t)$ with $\tilde{j}(t<0)=0 \;; 
\; \tilde{j}(t>0) \neq 0$ and using the property (\ref{zeroint}) the
equation of motion for the velocity change becomes
\begin{equation}
 M \dot{v} (t) + \int_{0}^t {\rm d} t^\prime \,
\Sigma_m(t-t^\prime)\,v(t^\prime) = \tilde{j}(t).\label{eqnofmotlap}
\end{equation}

The solution of this equation is found by Laplace transform,
in terms of $\tilde{v}(s)\; ; \;  \tilde {\Sigma}_m(s)\; ; \; \tilde{J}(s)$,
the Laplace transforms of the velocity, self-energy kernel and current
respectively, in terms of the Laplace variable $s$. We find that the
solution is given by
\begin{equation}
\tilde{v}(s) = \frac{v_0+(\tilde{J}(s)/M)}{s+\frac{1}{M}\tilde{\Sigma}_m(s)}.
\label{laplasol}
\end{equation}
The quantity 
\begin{equation}
G(s) = \frac{1}{s+\frac{1}{M}\tilde{\Sigma}_m(s)}
\label{propag}
\end{equation}
is the Laplace transform of the propagator of the velocity of the collective 
coordinate. We can now extract the frequency dependent conductivity associated
with the moving soliton by taking $v_0=0$ and analytically continuing 
$s \rightarrow i\omega+0^+$ to obtain the retarded Fourier transform. 
We find
\begin{equation}
\sigma(\omega) = \frac{\Delta\phi(e {\cal C})^2}{M}G(s=i\omega+0^+)
\label{conductivity}
\end{equation}
Therefore the frequency dependent conductivity is solely determined by $G(s)$ 
which can be found in a consistent adiabatic expansion. 

The real time evolution is found by the inverse Laplace transform
\begin{equation}
v(t) = \frac{1}{2\pi i} \int_C e^{st} \tilde{v}(s) ds
\end{equation}
where $C$ refers to the Bromwich contour running along the imaginary
axis to the right of all the singularities of $\tilde{v}(s)$ in the
complex $s$ plane. Therefore we need to understand the analytic 
structure  of $G(s)$ in eqn. (\ref{laplasol}) to obtain the real time dynamics. 
The Laplace transform of the self-energy kernel is given by
\begin{eqnarray}
\tilde{\Sigma}_m(s) & = & s \tilde{\Gamma}_m(s) \label{lapfric} \\
\tilde{\Gamma}_m(s)       & = & -4\,  \sum_{p,k\not=0} \left\{ T_{pk}^{(S)}\, 
T_{-p-k}^{(S)} (1 \,+ 2 n_p) \,\frac{(\omega_p + \omega_k)}{s^2+(\omega_p + 
\omega_k)^2} \right. \nonumber \\
&& \left. -  \quad 2 T_{pk}^{(A)} \, T_{-p-k}^{(A)} n_p 
\frac{(\omega_p - \omega_k)}{s^2+(\omega_p - \omega_k)^2}
\right\}, \label{laps}
\end{eqnarray}
 where $\tilde{\Gamma}_m(s)$ is the Laplace transform of the kernel
$\Gamma_m$ given above. 

The presence of a static friction coefficient will be revealed by a
pole in $G(s)$ with a negative real part, since this will translate into 
an exponential relaxation of the velocity.

In the absence of
interactions $G(s)$ has a simple pole at $s=0$. Since we obtained the
expression for the kernels in perturbation theory the position of a
pole must be found in a consistent perturbative expansion by writing
$s_p= (1/M) s_1 + \cdots $, we find
\begin{equation}
s_p = \tilde{\Sigma}_m(s=0) \equiv 0. \label{staticfric}
\end{equation} 
 Therefore the coefficient of {\em static} friction vanishes.
This is a consequence of the vanishing of the integral (\ref{zeroint}). 
Therefore up to this order in perturbation theory the position of the
pole in the s-variable remains at $s=0$. This is consistent with the
results of Ogata et. al.\cite{ogata} who also found that to lowest order
in the adiabatic expansion the {\em static} friction coefficient vanishes.

From the expression (\ref{laps}) we also find that $G(s)$ has cuts along
the imaginary s-axis: i) a two phonon cut beginning at $s= \pm 2i \omega_o$
corresponding to the virtual processes of spontaneous and stimulated two-phonon
creation and
destruction, ii) a cut with a pinch singularity beginning at $s= i 0^{\pm}$ 
corresponding to the processes of phonon scattering. The contribution from this
second cut vanishes at $T=0$. In summary, the analytic structure of $G(s)$ in
the complex 
s-plane corresponds to a pole at $s=0$ with residue
\begin{equation}
Z_s = \frac{1}{1+\frac{1}{M}\tilde{\Gamma}_m(0)} \label{wavefunc}
\end{equation} 
and cuts along the imaginary axis beginning at $\pm 2i \omega_o \; ; 
\pm i\epsilon$ 
with $\epsilon \rightarrow 0$ to clarify that the beginning of this cut
pinches the pole.

The residue $Z_s$ has a very clear interpretation, it is the  
``wave function renormalization'' and its effect can be understood in
two alternative manners.

Consider the case in which $\tilde{j}=0$ in eqn.(\ref{laplasol}). 
Performing the inverse Laplace transform and invoking the Riemann-Lebesgue lemma,
the long time behavior will be completely dominated by the pole at $s=0$. 
Therefore, if the velocity of the soliton has been changed at $t=0$ by some
external source, this disturbance will relax in time to an asymptotic value
given by
\begin{equation}
v_{\infty} = Z_s v_0. \label{asinto}
\end{equation}

Alternatively, consider the case of $v_0=0$ but with an electric field switched
on at $t=0$ and constant in time thereafter. Again the inverse Laplace transform
at long time will be dominated by the pole,  and we find that the soliton moves
with constant acceleration given by
\begin{equation}
\dot{v}  = \frac{\tilde{j}}{M_{eff}} \quad \mbox{ with } \quad 
M_{eff}  =  \frac{M}{Z_s}. 
\label{acceleration} 
\end{equation}
Thus the wave function renormalization can also be understood as a renormalization 
of the soliton mass. The ratio of the asymptotic acceleration to the
initial acceleration is given by $Z_s$. As the soliton moves, the
interaction with the phonon (phason) bath ``dress'' it changing its 
effective mass, which will be seen in specific models  to be {\em larger} 
than the bare mass.  This result is similar to that  found by Holstein and
Turkevich in the polaron case within a different approach\cite{c6}. 

A further understanding of the dynamics will necessarily require knowledge of
the matrix elements to establish the details of the kernels. This will be studied
in particular models in the next section.

\subsection{Semiclassical Langevin Equation}

The classical Langevin equation is an adequate phenomenological description of
Brownian motion obtained by considering the dynamics of one (or few) degrees 
of freedom that interact with a bath in equilibrium. 

It contains a term proportional to the velocity of the particle which 
incorporates friction and dissipation and a stochastic term which reflects 
the random interaction of the heat bath with the particle. These two terms 
are related by the classical fluctuation-dissipation relation. 

At the quantum mechanical level it is also possible to obtain  a ``reduced'' 
or coarse grained description of the dynamics of one (or few) degrees of freedom
in interaction with a bath. The coarse graining procedure has a very precise
meaning: the full time dependent density matrix is traced over the bath degrees
of freedom yielding an effective or ``reduced'' density matrix for the degrees
of freedom whose dynamics is studied.

 Such a  description of non-equilibrium dynamics of a quantum mechanical 
particle coupled to a dissipative environment by a Langevin equation was 
presented by Caldeira and Leggett\cite{caldeira} and by Schmid \cite{d6}. 
Their technique is based on the influence-functional method of Feynman and
Vernon \cite{d5} that naturally leads to a semiclassical  Langevin equation.

In this section we follow the procedure of\cite{caldeira}-\cite{d8} 
generalized to our case to derive the Langevin equation for solitons in a 
heat bath to lowest order in the adiabatic coupling. 

The main step is to perform the path integrals over the phonon degrees of
freedom, thus obtaining an effective functional for the collective coordinate
of the soliton. Unlike the most usually studied cases of a particle linearly 
coupled to an harmonic reservoir\cite{caldeira}-\cite{d8} we have here a 
bilinear coupling to the phonons (phasons). Therefore the influence functional
cannot be obtained exactly,
but it can be obtained in a consistent perturbative expansion. For this we
treat  the interaction term ${\cal L}_I[\dot{x}^{\pm},\gamma^{\pm},\gamma^{*\pm}]$
in perturbation theory up to second order in the vertex proportional to 
$\dot{x}^{\pm}$ (which is equivalent to lowest order
in the adiabatic coupling $m/M$). Integrating over the phonon variables and
using $<F[a^\dagger,a]>=0$, we obtain
\begin{equation}
{\cal Z}[j^+,j^-=0] \,=\, \int {\cal D}x^+\,{\cal D}x^-\, \;
e^{i\int_{-\infty}^{\infty} dt^\prime \left({\cal L}_0[\dot{x}^+]\,-\,
{\cal L}_0[\dot{x}^-]\right)} \;{\cal F}[\dot{x}^+,\dot{x}^-],
\label{infl}
\end{equation}
where
\begin{equation}
{\cal L}_0[\dot{x}^\pm] \,=\, \frac{1}{2} M \left(\dot{x}^\pm\right)^2 - \tilde{j}
x^{\pm}
\label{L0}
\end{equation}
and ${\cal F}[\dot{x}^+,\dot{x}^-]$ is the influence 
functional\cite{d5}-\cite{d8}. To
lowest adiabatic order we find
\begin{eqnarray}
{\cal F}[\dot{x}^+,\dot{x}^-] & = & exp\left\{-\, \frac{1}{2} \int dt\, 
dt^\prime \right. \biggl[
\dot{x}^+(t) \, G^{++}(t,t^\prime) \, \dot{x}^+(t^\prime) \,+\, \dot{x}^-(t) 
\, G^{--}(t,t^\prime) \, \dot{x}^-(t^\prime) \nonumber \\
&  & \hspace{1in} + \; \left.  \dot{x}^+(t) \,G^{+-}(t,t^\prime) \,
\dot{x}^-(t^\prime) \,+\,
\dot{x}^-(t) \, G^{-+}(t,t^\prime) \, \dot{x}^+(t^\prime) \biggr] \right\}
\end{eqnarray}
in terms of the real-time phonon correlation functions  (see appendix (\ref{B}))
\begin{eqnarray}
G^{++}(t,t^\prime) & = &  <F[a^{\dagger +}(t),a^+(t)]\;F[a^{\dagger +}(t^\prime)
,a^+(t^\prime)]> \nonumber \\
G^{--}(t,t^\prime) & = &  <F[a^{\dagger -}(t),a^-(t)]\;F[a^{\dagger -}(t^\prime)
,a^-(t^\prime)]>\nonumber \\
G^{+-}(t,t^\prime) & = &  -\,<F[a^{\dagger +}(t),a^+(t)]\;F[a^{\dagger -}
(t^\prime),a^-(t^\prime)]>\nonumber \\
G^{-+}(t,t^\prime) & = &  -\,<F[a^{\dagger -}(t),a^-(t)]\;F[a^{\dagger +}
(t^\prime),a^+(t^\prime)]>.
\label{greens}
\end{eqnarray}

At this stage it is convenient to introduce the center of mass and relative 
coordinates, $x$ and $R$ respectively, which are defined as
\begin{equation}
x(t)\,=\, \frac{1}{2} \left( x^+(t)\,+\,x^-(t) \right) \hspace{.3in},\hspace{.3in}
R(t)\,=\,x^+(t)\,-\,x^-(t).
\label{wigner}
\end{equation}
These are recognized as the coordinates used in the Wigner transform of 
the density matrix \cite{caldeira}-\cite{d8} in terms of which the partition
function becomes
\begin{equation}
{\cal Z}[0] \,=\, \int {\cal D}x\,{\cal D}R\, \;e^{iS[x,R]}
\label{p2}
\end{equation}
with the non-equilibrium effective action given by 
\begin{equation}
S[x,R] \,=\, \int dt \, R(t) \, \left[ -M \ddot{x}(t) \,-\, \frac{i}{2}\int 
dt^\prime\,\biggl(
K_1(t-t^\prime)\, \dot{x}(t^\prime) \,-\,K(t-t^\prime)\,R(t^\prime) \biggr) 
\right] 
\label{action}
\end{equation}
in terms of  the kernels $K_1(t-t^\prime)$ and $K(t-t^\prime)$ which are given 
by (see appendix (\ref{C}) )
\begin{eqnarray}
K_1(t-t^\prime) & = & 8\,i\, \theta(t-t^\prime) \sum_{p,k\not=0} 
\biggl\{ T_{pk}^{(S)} \,
T_{-p-k}^{(S)} (1 +  n_p + n_k ) (\omega_p + \omega_k) 
\cos \left[ (\omega_p + \omega_k)(t - t^\prime)\right]    \nonumber \\
&& - \;T_{pk}^{(A)}\,T_{-p-k}^{(A)}( n_p - n_k )\,
(\omega_p - \omega_k) \cos\left[(\omega_p - \omega_k)
(t - t^\prime)\right] \biggr\} \nonumber \\
& = & -2\,i\,\Sigma_m(t-t^\prime)
\label{ker1}
\end{eqnarray}
and
\begin{eqnarray} 
K(t-t^\prime) & = & -2\, \sum_{p,k\not=0} \biggl\{ T_{pk}^{(S)} \,
T_{-p-k}^{(S)} (1 +  n_p + n_k + n_p\,n_k) (\omega_p + \omega_k)^2 
\cos \left[ (\omega_p + \omega_k)(t - t^\prime)\right]  
 \nonumber \\
&& + \,2\, \;T_{pk}^{(A)}\,T_{-p-k}^{(A)} n_k( 1 + n_p )\,
(\omega_p - \omega_k)^2 \cos\left[(\omega_p - \omega_k)
(t - t^\prime)\right] \biggr\}. \label{ker}
\end{eqnarray}

At this stage it proves convenient to introduce the identity 
\begin{eqnarray}
e^{-\frac{1}{2}\int dt \, dt^\prime R(t) \; K(t-t^\prime)\, R(t^\prime)} & = & 
C(t) \int {\cal D} \xi \;e^{-\frac{1}{2} \int dt \, dt^\prime  \xi(t)K^{-1}
(t-t^\prime)\xi(t^\prime) \,+\, i\int dt\,\xi(t) R(t) }
\end{eqnarray}
with C(t) being an inessential normalization factor, to cast the 
 non-equilibrium effective action of the collective coordinate
in terms of a stochastic noise variable with a definite probability
 distribution\cite{d6}-\cite{d8}.

\begin{eqnarray}
{\cal Z}[0] &=& \int {\cal D}x\,{\cal D}R\,{\cal D} \xi \,P[\xi]\; 
exp\biggl\{i \int dt R(t) \biggl[ -M \ddot{x}(t) \,-\, \frac{i}{2}\int dt^\prime\,
K_1(t-t^\prime)\, \dot{x}(t^\prime) \,+\, \xi(t) \biggr] \biggr\}, \nonumber \\
\label{ZLang}
\end{eqnarray}
where the probability distribution of the stochastic noise, $P[\xi]$, is given by
\begin{equation}
P[\xi]\,=\,\int {\cal D} \xi \;exp\left\{ -\frac{1}{2} \int dt \, dt^\prime  
\xi(t)K^{-1}(t-t^\prime)\xi(t^\prime) \right\}.
\end{equation}
In this approximation we find that the noise is Gaussian, additive and with
correlation function  given by
\begin{equation}
< \xi(t) \xi(t^\prime)>\,=\, K(t-t^\prime).
\end{equation}

The semiclassical Langevin equation is obtained by extremizing  the effective 
action  in eqn.(\ref{ZLang})with respect to $R(t)$\cite{caldeira}-\cite{d8}
\begin{equation}
M \ddot{x}(t) \,+\, \int_{-\infty}^t dt^\prime\,
\Sigma_m(t-t^\prime)\, \dot{x}(t^\prime)-\tilde{j}(t) \,=\, \xi(t).
\label{Lang}
\end{equation}

Two features of the semiclassical Langevin equation  deserve comment. The first 
is that the kernel $K_1(t-t^\prime)$, as can be seen from eqn.(\ref{ker1}), 
is non-Markovian. The second is that the noise correlation function 
$K(t-t^\prime)$ is colored, i.e. it is not a delta function 
$\delta(t-t^\prime)$. The relationship between the kernels $K_1(t-t^\prime)$
and $K(t-t^\prime)$ established in appendix (\ref{C}) constitutes a 
generalized quantum fluctuation dissipation 
relation\cite{caldeira}-\cite{d8}.
Finally we recognize that taking the average of (\ref{Lang}) with the
noise probability distribution $P[\xi]$ yields the equation of motion for
the expectation value of the collective coordinate (eqn.(\ref{eqmotm})).

A classical description is expected to emerge when the occupation distribution for
the phonons can be approximated by their classical counterparts\cite{caldeira},
i.e. when $n_k \approx T/\omega_k$ (in units in which the natural constants had
been set to 1).
 However, this classical limit requires that $T \gg m$ and with $m$ being
identified with the optical phonon  or phason frequency in these models, such a
classical approximation will be valid when temperatures are much larger than 
these  frequencies.
In the models under consideration the optical phonon 
frequencies are in the range $\omega_o \approx 0.1 \mbox{ eV }$ and the phason
frequencies (in the case of charge density wave systems)$\omega_o \approx 
10^{-3} \mbox{ eV }$. A more stringent criterion for the validity of the
classical limit is when the temperature is larger than the 
band-width\cite{caldeira}. In the situations under consideration the 
band-width is typically several eV. Hence a classical
description will be valid in a temperature regime that far exceeds the
experimentally relevant region in the case of conducting polymers. In the case 
of charge density wave systems the experimentally relevant temperatures are
of the order of a few times $\omega_o$. For these systems whether the fluctuation
and dissipation kernels achieve a classical limit for these temperatures must
be studied in detail. This will be done with  particular model Hamiltonians below. 

{\em If} the kernels $\Sigma_m$ and $K$ admit a Markovian limit then a diffusion 
coefficient could be extracted by computing the long time limit of the 
correlation function $<<(x(t)-x(0))^2>>/t$ where $<<\cdots>>$ stand for 
average over the noise distribution function. However when the kernels do
not become Markovian, such a definition is not appropriate.  

This summarizes the general formulation of the description of the dynamics
of the collective coordinate both at the level of the evolution equation for
the expectation value as well as for the effective Langevin dynamics in terms
of stochastic noise terms arising from the fluctuations in the phonon bath.
We are now in condition to study specific models.

\section{ Specific Models}

In the previous sections we established the general aspects of the real-time
dynamics of solitons in the presence of the phonon bath, obtaining the 
equation of motion as well as the Langevin equation for the collective coordinate
in lowest adiabatic order. Further progress in the understanding of the dynamics
necessarily involves the details of particular models which determine the
matrix elements $T^{(A,S)}$ and therefore the time dependence of the kernels
involved. In this section we study these details for the Sine-Gordon
and $\phi^4$ models. 

\subsection{ Sine-Gordon }
As discussed in section 2, Sine-Gordon field theory provides an effective
microscopic description for phase solitons in CDW systems\cite{d3}-\cite{horo} 
in the
limit in which the amplitude of the lattice distortion is kept constant. 

For the Sine-Gordon model the potential is given by 
\begin{equation}
U(\phi) = {m^4\over g} \left( 1 - \cos\left[ {\sqrt g \over m} 
\phi \right] \right)
\label{eigfsg}
\end{equation}
and the static soliton solution is given by\cite{d2,c7,c51}
\begin{equation}
\phi_s(x) = {4 m \over{\sqrt g}} {\rm arctan} \left[e^{  mx}\right]\,.
\label{phissg}
\end{equation}

The normal modes of this theory are the solutions of the following equation,
(see eqn.(\ref{hosc})) 
\begin{equation}
\left[ - \frac{d^2}{dx^2} \,+ \,m^2\,-\,\frac{2 m^2}{cosh^2(mx)}\, \right]
\,\psi_n(x)\,=\,\omega^2_n
\psi_n(x). 
\end{equation}
The solutions of the above differential equation are well known \cite{d1},
\cite{d2},\cite{c51}. There is only one bound state with zero eigenvalue, 
the zero mode, followed by a continuum with wave functions given by
\begin{equation}
{\cal U}_k(x) = {1\over{\sqrt{2 \pi}\omega_k}}(-ik \,+ \, m \, {\rm tanh} 
(m\,x))e^{i k x} 
\label{Usg}
\end{equation}
with $\omega_k^2 = k^2 + m^2$. The scattering states represent the phason
(harmonic fluctuations of the phase) excitations around the 
soliton\cite{d3}-\cite{horo}.

The matrix elements $D_{pk}$ were already calculated by de Vega\cite{c51}, 
(see eqns.(\ref{Dmn},\ref{me1})) and are given by
\begin{eqnarray}
D_{kp} \,=\, i p \delta(k + p) + {i ( p^2 - k^2) \over{ 4 \omega_k \omega_p 
\sinh\left[ {\pi \over 2} {(p + k)\over m}\right]}} \quad \hbox{ for }
 p \neq k  
\label{elmatsg}
\end{eqnarray}
which determine the symmetric and antisymmetric matrix elements 
\begin{eqnarray}
T_{pq}^{(S)} &=& {1\over 4} \left[ \left({\omega_p\over{\omega_q}}
\right)^{1/2} - \left({\omega_q\over{\omega_p}}\right)^{1/2}\right] 
\left\{ {( q^2 - p^2) \over{ 4 \omega_k \omega_p 
\sinh\left[ {\pi \over 2} {(q + p)\over m}\right]}} \right\} \nonumber \\
T_{pq}^{(A)} &=& {1\over 4} \left[ \left({\omega_p\over{\omega_q}}
\right)^{1/2} + \left({\omega_q\over{\omega_p}}\right)^{1/2}\right] 
\left\{ {( q^2 - p^2) \over{ 4 \omega_k \omega_p 
\sinh\left[ {\pi \over 2} {(q + p)\over m}\right]}} \right\} \,.
\label{TaTssg}
\end{eqnarray}

Since in this theory there are no bound states other than the zero mode, 
$F[a^\dagger,a]$ is given only by the first two terms in eqn.(\ref{foobd}).
Substituting eqn.(\ref{TaTssg}) in eqns.(\ref{GTT}) and eqn.({\ref{sigmf}), 
we obtain the final form of the kernels in this case
\begin{eqnarray}
\Gamma_m(t - t^\prime) &=& {1\over{4^3}} \int_{-\infty}^\infty 
{{\rm d} p \, {\rm d} k \over{ \omega_p^3 \omega_k^3}} 
{(p^2 - k^2)^2 \over{\sinh^2\left[ 
{\pi \over 2} {(p + k)\over m}\right]}} \Biggl\{ ( 1 + 2n_p) (\omega_p - 
\omega_{k})^2 \sin\left[(\omega_p + \omega_{k})
(t - t^\prime) \right]   \nonumber \\
&& \qquad \qquad  - 2\quad n_p  (\omega_p + \omega_{k})^2 
\sin\left[(\omega_p - \omega_{k})(t - t^\prime)\right] \Biggr\} \label{GTTM}
\end{eqnarray}
\begin{eqnarray}
\Sigma_m(t - t^\prime) &=& {1\over{4^3}} \int_{-\infty}^\infty 
{{\rm d} p \, {\rm d} k \over{ \omega_p^3 \omega_k^3}} 
{(p^2 - k^2)^3 \over{\sinh^2\left[ 
{\pi \over 2} {(p + k)\over m}\right]}} \Biggl\{ (\omega_p - 
\omega_{k})( 1 + n_p + n_k) \cos\left[(\omega_p + \omega_{k})
(t - t^\prime) \right]   \nonumber \\
&& \qquad \qquad  - \quad (n_p - n_k)(\omega_p + \omega_{k}) 
\cos\left[(\omega_p - \omega_{k})(t - t^\prime)\right] \Biggr\}.
\label{sigmsg}
\end{eqnarray} 

At this point it proves useful to 
express $\Gamma_m(t - t^\prime)$ and $\Sigma_m(t - t^\prime)$ in terms of 
dimensionless quantities to display at once the nature of the
adiabatic expansion. To achieve this let us make the following change of 
variables
\begin{equation}
p  \rightarrow \frac{p}{m}\quad ;\quad k  \rightarrow \frac{k}{m}\quad ; \quad 
\tau = m\,t\quad \hbox{ and } \quad {\cal T} = {T\over  m} \label{change} \, .
\end{equation}
Then $\Gamma_m(t - t^\prime)$ and $\Sigma_m(t - t^\prime)$ can be written as
\begin{equation}
\Gamma_m(t - t^\prime) \,=\, m^2\; \Gamma(\tau - \tau^\prime)\quad\quad
\hbox{ and } \quad\quad
\Sigma_m(t - t^\prime) \,=\, m^3\; \Sigma(\tau - \tau^\prime),
\label{dimless}
\end{equation}
where
\begin{eqnarray}
\Gamma(\tau)&=&\int_{-\infty}^\infty {\rm d} p\,{\rm d} k \left\{\Gamma_1(p,k)
\,\sin\left[(w_p + w_{k})(\tau ) \right]\;+\;\Gamma_2(p,k)\,\sin\left[
(w_p - w_{k})(\tau ) \right]\right\} \label{GTTMm} \\
\Sigma(\tau)&=&\int_{-\infty}^\infty {\rm d} p\,{\rm d} k \left\{\Sigma_1(p,k)
\,\cos\left[(w_p + w_{k})(\tau) \right]\;+\;\Sigma_2(p,k)\,\cos\left[
(w_p - w_{k})(\tau) \right] \right\}
\label{sigsg}
\end{eqnarray}
with
\begin{eqnarray}
\Gamma_1(p,k) &=& {1\over{32}}\;{( 1 + 2n_p ) (p^2 - k^2)^2 (w_p\,-\,w_k)^2 
\over{w_p^3\,w_k^3\,\sinh^2\left[ {\pi \over 2} (p + k)\right]}} \nonumber \\
\Gamma_2(p,k) &=& - {1\over{16}}\;{n_p  (p^2 - k^2)^2 (w_p\,+\,w_k)^2 
\over{w_p^3\,w_k^3\,\sinh^2\left[ {\pi \over 2} (p + k)\right]}} \nonumber \\
\Sigma_1(p,k) & = & (w_p\,+\,w_k)\;\Gamma_1(p,k) \nonumber \\
\Sigma_2(p,k) & = & (w_p\,-\,w_k)\;\Gamma_2(p,k) \nonumber \\
w_p^2  & = & p^2 +1 \; ; \; 
n_p  =  \frac{1}{e^{\frac{ w_p}{ {\cal T}}}-1}. 
\end{eqnarray}

Fig.(\ref{figssg}) shows the numerical evaluation of $\Gamma(\tau)$ and 
$\Sigma(\tau)$ vs. $\tau$ for different values of ${\cal T}$. We clearly see
that the self-energy kernel $\Sigma$ is peaked near $\tau =0$ and localized
within a time scale $\tau_s \approx m^{-1} \approx \omega_0^{-1}$ at low 
and intermediate temperatures. We find numerically that this time scale becomes
very short, of the order of $T^{-1}$ for $T \geq 10 m$ which for the
case of charge density wave systems is about the maximum temperature scale of
experimental relevance. Similarly, the kernel $\Gamma$ varies slowly over a 
large time scale $ \approx 5-10 m^{-1} $ for large temperatures, but at 
small and intermediate temperatures $T \leq m  \;(\omega_o)$ it oscillates
within time scales comparable to the inverse phason frequency.

\subsubsection{Equation of Motion: i) exact solution}

With the purpose of providing a numerical solution to the equations of motion,
we now consider the case  of an externally applied electric field switched on
at $t=0$ and maintained constant in time thereafter. In terms of dimensionless
quantities the equation of motion (\ref{eqmotm}) becomes in this case

\begin{equation}
\dot{v} (\tau) + \frac{m}{M}\int_0^\tau {\rm d} \tau^\prime \,
\Sigma(\tau-\tau^\prime)v(\tau^\prime) = j,
\label{seem}
\end{equation}
where $j = \tilde{j}/(mM) \; $and the dot stands for derivative with respect to
the dimensionless variable $\tau$. 

We will choose the initial condition $v_0=v(\tau=0)=0$. From the solution
$v_j(\tau)$ of eqn. (\ref{seem}) with this initial condition, the solution to
the homogeneous equation with $v_0 \neq 0$ is obtained as
\begin{equation}
v(\tau) = v_0 \frac{\dot{v}_j(\tau)}{j} \label{vosol}
\end{equation}

\noindent and the general solution is given by the sum of the inhomogeneous and
homogeneous ones. 

\subsubsection{Equation of Motion: ii) the Markovian approximation}

As shown in fig.(\ref{figssg}), the kernel $\Sigma(\tau)$  has ``memory'' 
on time scales a few times the inverse of the phason frequency at low and
intermediate temperatures $T \leq m$. If the soliton velocity varies on 
time scales larger than the ``memory'' of the kernel  a Markovian 
approximation to the dynamics may be reasonable. The first step in the 
Markovian approximation corresponds to replacing $v(\tau')$ by $v(\tau)$
inside the integral in eqn. (\ref{seem}) and taking it outside the integral.
A second stage of approximation would take the upper limit of the integral 
to $\infty$ thus integrating the peak of the kernel. However we have shown 
above that the total integral of the kernel vanishes, thus this second stage
cannot be invoked. Recognizing that
$\int_0^{\tau} \Sigma(\tau -\tau')d\tau' = \Gamma(\tau)$ the Markovian 
approximation to (\ref{seem}) is given by

\begin{equation}
\dot{v}(\tau) + \frac{m}{M}v(\tau)\Gamma(\tau)=j.
\label{smem}
\end{equation}

As advanced in the previous section, we now identify the kernel $\Gamma(\tau)$
as the {\em dynamical} friction coefficient in the Markovian approximation.
The property (\ref{zeroint}) determines that $\Gamma(\tau \rightarrow \infty) =0$.

Figure (\ref{figsfv}) shows numerical solutions of eqn.(\ref{seem}) and 
eqn.(\ref{smem}) for temperatures ${\cal T}$ = 0, 1.0, 5.0 and 10.0. As can 
be seen from the figures,  the departure of the exact solution from a straight
\footnote{We refer to the numerical solution of eqn.(\ref{seem}) as the 
exact solution to distinguish it from the numerical solution in the Markovian
approximation, eqn.(\ref{smem})}
line (free case) is larger the larger the ratio $m/M$ and the temperature 
${\cal T}$. This is expected since  larger  adiabatic ratio implies a stronger
coupling between soliton and bath, whereas  larger the temperature implies that 
 more phonons are excited in the bath that contribute to the scattering
term and stimulated  creation and absorption of excitations.  
At zero temperature, the soliton moves experiencing negligible dissipative
force, since to dissipate energy the soliton needs to excite two phonons in
a virtual
state, but there is a gap for this process making it rather inefficient.
The solution in  the Markovian approximation, $v_m(\tau)$, is almost 
indistinguishable from the free evolution even
at very large temperature and couplings.
Thus we see that memory effects
are extremely important even at high temperatures and a Markovian approximation
will be unwarranted at least to the order in which this calculation has been
performed.

\subsubsection{Velocity relaxation and wave function renormalization }

In order to display more clearly the dissipative effects, we now study the
relaxation of the soliton velocity. For this consider that $j(t>0) =0$ but  an
initial velocity $v_0$ at $t=0$. With this initial condition and $j \equiv 0$,
eqn. (\ref{seem}) becomes an initial value problem.

As the soliton moves in the bath, its velocity decreases because of 
the interaction with the fluctuations,  the asymptotic final velocity is 
related to
the initial velocity through the wave function renormalization as explained 
section 4.2 above. We  present the numerical solution of the homogeneous
equation with initial velocity $v_0=1$ in figure 3, where we also present the
homogeneous solution in the Markovian approximation described above. We 
clearly see that the initial velocity relaxes to an asymptotic value $v_{\infty}$.
However the time dependence cannot be fit with an exponential. 

According to the analysis of the general solution, 
the ratio $v_{\infty}/ v_0$ should be given by the wave function renormalization, 
i.e.
\begin{equation}
Z_s = \frac{1}{1+\frac{m}{M}\tilde{\Gamma}(s=0)}= \frac{v_{\infty}}{v_0}.
\label{wavefc}
\end{equation}

Table 1 below compares the ratio $v_{\infty}/v_0$ obtained from the numerical
solution to the exact evolution equation, with the value of the wave-function 
renormalization. Clearly the agreement is excellent, confirming the analysis 
of the asymptotic behavior of the solution in real time.

\begin{center}
\begin{tabular}{||c||c|c||c|c||} 
\multicolumn{5}{c}{ \textbf{Table (1)} } \\ 
\multicolumn{5}{c}{  } \\ \hline \hline
   & \multicolumn{2}{c||}{ $v_\infty / v_0$ } & \multicolumn{2}{c||}{ $Z_s$ }
   \\ \hline
   & m/M = 0.1 & m/M = 0.25& m/M = 0.1 & m/M = 0.25 \\  \hline\hline
Zero Temp. &  0.999808 & 0.999521 & 0.999808 & 0.999521 \\ \hline
Temp. 1.0 &  0.993438 & 0.983754 & 0.993438 &  0.983753 \\ \hline
Temp. 5.0 &  0.96055 & 0.906885 &  0.96055 & 0.90687 \\ \hline
Temp. 10.0 & 0.923458 & 0.828352 & 0.923446 & 0.828303 \\ \hline \hline 
\end{tabular}
\end{center}

\subsubsection{ Kernels for the  semiclassical Langevin equation}

Knowledge of the matrix elements $T^{(A)},T^{(S)}$ allow us to obtain
the final form of the kernels that enter in the semiclassical Langevin 
equation given by eqns.(\ref{ker1}) and (\ref{ker}), and eqn.(\ref{TaTssg}).
These kernels can be written in terms of the dimensionless quantities given
by eqn.(\ref{change}). 
Since $K_1(t-t') = -2i\Sigma_m(t-t')$ we focus on $K(t-t')$.
In term of dimensionless quantities, $K(t)= m^4 {\cal K}(\tau)$ where

\begin{equation}
{\cal K}(\tau)\,=\, \int_{-\infty}^\infty {\rm d} p\,{\rm d} k 
\left\{C_1(p,k)\,\cos\left[(w_p + w_{k})(\tau) \right]\;+\;C_2(p,k)\,
\cos\left[(w_p - w_{k})(\tau) \right] \right\}.
\label{kersingor}
\end{equation}
with
\begin{eqnarray}
C_1(p,k) &=& {2\over{4^4}}\;{( 1 + n_p + n_k + n_p\,n_k) (p^2 - k^2)^4  
\over{w_p^3\,w_k^3\,\sinh^2\left[ {\pi \over 2} (p + k)\right]}} 
\label{c1singor} \\
C_2(p,k) &=& {1\over{4^3}}\;{n_k(1 + n_p) (p^2 - k^2)^4  
\over{w_p^3\,w_k^3\,\sinh^2\left[ {\pi \over 2} (p + k)\right]}} 
\label{c2singor}
\end{eqnarray}

Fig.(\ref{figscr}) shows ${\cal K}(\tau)$ for different temperatures ${\cal T}$.
Notice that at large temperatures the kernel becomes strongly peaked at
$\tau =0$ and one would be tempted to conclude that the classical limit
corresponds to a delta function. However the coefficients 
(\ref{c1singor},\ref{c2singor}) are such that the total integral in 
$\tau$ (leading to delta functions of sums and differences of frequencies) 
vanishes. We then conclude that even in the high temperature limit the 
noise-noise correlation function is not a delta function, i.e. the noise is
``colored'', the classical fluctuation dissipation relation does not emerge 
and  a diffusion coefficient cannot
be appropriately defined.

\subsection{$\phi^4$ Theory}

In this model, originally studied by Krumanshl and Schrieffer\cite{krum}, 
the interaction is given by
\begin{equation}
U(g,\phi) = {1\over {2g}} \left( m^2 - g \phi^2 \right)^2,
\label{uphi}
\end{equation}
where $m$ is a parameter with dimension of mass. The static soliton solution
is given by
\begin{equation}
\phi_s(x-x_0) =  { m \over{\sqrt g}} {\rm tanh} \left[m(x-x_0)\right]\,,
\label{phifis}
\end{equation}

\noindent and the normal modes are the solutions to the equation, 
(see eqn.(\ref{hosc}))
\begin{equation}
\left[ - \frac{d^2}{dx^2} \,+ \,4 m^2\,-\,\frac{6 m^2}{{\rm cosh}^2(mx)}\,
\right]\,\psi_n(x)\,=\,\omega^2_n
\psi_n(x). 
\end{equation}
The solution of the above differential equation is well known \cite{d1},
\cite{d2}. It has two bound states followed by a continuum. The normalized
eigenvectors are given by
\begin{eqnarray}
{\cal U}_0(x) &=& \frac{ \sqrt{3\,m} }{2} {\rm sech}^2[m\,x] \propto 
\frac{d \phi_s}{dx} \quad\quad \mbox{with} \quad\quad
\omega_0\,=\,0 \nonumber \\
{\cal U}_b(x) &=& \frac{ \sqrt{3\,m} }{2} \, {\rm sech}[m\,x] \, 
{\rm tanh}[m\,x]\quad\quad \mbox{with} \quad\quad \omega_b^2\,=\,3\,m^2 
\nonumber \\
{\cal U}_k(x) &=& {m^2 e^{i k x}\over{\sqrt{2\pi (k^2 + m^2)} \omega_k}}
\left\{ 3 {\rm tanh}^2[m x] - 3 \, i \,{k\over m}\, {\rm tanh} [m x] - 1 -
\frac{k^2}{m^2} \right\} 
\label{Up4}
\end{eqnarray}
with $\omega_k^2 = k^2 + 4\,m^2$. The scattering states are identified with
optical phonon modes and the optical phonon frequency is identified with 
$\omega_o=2m$. 

The bound state with zero frequency is the ``zero mode'', whereas
the bound state with $\omega^2_b = 3m^2$ corresponds to an amplitude 
distortion\cite{krum,c7} of the soliton. 

The matrix elements $D_{pk}$  are given by, (see eqns.(\ref{Dmn},\ref{me1}))

\begin{eqnarray}
D_{bk} &=& \frac{\sqrt{3 \pi}}{8}\,\frac{ {\rm sech}\left[ \frac{\pi k}{2\,m} \right] }{m^\frac{3}{2}\; \omega_k } \, \sqrt{k^2+m^2}\;(k^2+3\,m^2) \quad\quad \hbox{ (from the bound state) }\nonumber \\
D_{pk} &=& i k \delta(p + k) + {3 \,i \,  \pi ( k^2 - p^2) 
(p^2 + k^2 + 4 m^2)\over{ 4  m^4 N_p\,N_k
\sinh\left[ {\pi \over 2} {(p + k)\over m}\right]}} \quad \hbox{ for }
 p \neq k \; ,  
\label{elmatp4}
\end{eqnarray}
where $N_k$ is defined as
\begin{equation}
N_k = \sqrt{ {2 \pi w_k^2 (k^2 + m^2) \over{m^4}} }\,.
\end{equation}

We notice that the coupling to the continuum through the bound state given by 
the matrix element $D_{bk}$ is of the same order as the coupling to the 
continuum-continuum (matrix elements $D_{pk}$). This will have interesting
consequences  for the dissipational dynamics.  
The symmetric and antisymmetric matrix elements for the continuum states 
are  given by
\begin{eqnarray}
T_{pq}^{(S)} &=& {3\over 32} \left[ \left({\omega_p\over{\omega_q}}
\right)^{1/2} - \left({\omega_q\over{\omega_p}}\right)^{1/2}\right] 
\left\{ {( q^2 - p^2)(p^2 + q^2 + 4m^2)
\over{ \sqrt{q^2 + m^2} \sqrt{p^2 + m^2} \omega_q \omega_p 
\sinh\left[ {\pi \over 2} {(q + p)\over m}\right]}} \right\} \nonumber \\
T_{pq}^{(A)} &=& {3\over 32} \left[ \left({\omega_p\over{\omega_q}}
\right)^{1/2} + \left({\omega_q\over{\omega_p}}\right)^{1/2}\right] 
 \left\{ {( q^2 - p^2)(p^2 + q^2 + 4m^2)
\over{ \sqrt{q^2 + m^2} \sqrt{p^2 + m^2} \omega_q \omega_p 
\sinh\left[ {\pi \over 2} {(q + p)\over m}\right]}} \right\}\,.
\nonumber \\
&& \mbox{}
\label{TaTsp4}
\end{eqnarray}

\noindent whereas those involving the bound state are obtained by replacing 
the matrix elements $D_{bk}$ for the $D_{pk}$.

Since in this model there is one bound state other than the zero mode, the 
interaction vertex $F[a^\dagger,a]$ is given by eqn.(\ref{foobd}) in the 
appendix. 
The contributions from  bound-state-continuum virtual transitions do 
not mix with the continuum-continuum to this order in the adiabatic expansion.
As a consequence of this simplification the dimensionless kernels (in terms
of the dimensionless variables introduced in (\ref{change})) become

\begin{eqnarray}
\Gamma(\tau)&=& \int_{-\infty}^\infty {\rm d} p \biggl\{ \; \Gamma_1^b(p)\,
\sin\left[(w_p + w_{b})(\tau ) \right]\;+\;\Gamma_2^b(p)\,\sin\left[
(w_p - w_{b})(\tau ) \right] \nonumber \\
& & + \quad \int_{-\infty}^\infty
{\rm d} k \left\{\Gamma_1(p,k)\,\sin\left[(w_p + w_{k})(\tau ) \right]
\;+\;\Gamma_2(p,k)\,\sin\left[(w_p - w_{k})(\tau )\right]\right\} 
\;\biggr\}\nonumber \\
\Sigma(\tau)&=&\int_{-\infty}^\infty {\rm d} p \biggl\{ \; \Sigma_1^b(p)\,
\cos\left[(w_p + w_{b})(\tau ) \right]\;+\;\Sigma_2^b(p)\,\cos\left[
(w_p - w_{b})(\tau ) \right] \nonumber \\
& & + \quad \int_{-\infty}^\infty
{\rm d} k \left\{\Sigma_1(p,k)\,\cos\left[(w_p + w_{k})(\tau) \right]\;+
\;\Sigma_2(p,k)\,\cos\left[(w_p - w_{k})(\tau) \right] \right\}
\;\biggr\} \nonumber \\
\label{fisig}
\end{eqnarray}
with
\begin{eqnarray}
\Gamma_1(p,k) &\equiv& {3^2 \over{4^4}}\;{( 1 + n_p + n_k) (p^2 - k^2)^2 
(w_p\,-\,w_k)^2 \,
(p^2 + k^2 + 4)^2 \over{w_p^3\,w_k^3\,(p^2 + 1)\, (k^2+1) \sinh^2\left[ 
{\pi \over 2} (p + k)\right]}} \nonumber \\
\Gamma_2(p,k) &\equiv&  {3^2 \over{4^4}}\;{(n_k - n_p) (p^2 - k^2)^2 
(w_p\,+\,w_k)^2 \,
(p^2 + k^2 + 4)^2 \over{w_p^3\,w_k^3\,(p^2 + 1)\, (k^2+1)\,\sinh^2
\left[ {\pi \over 2} (p + k)\right]}} \nonumber \\
\Gamma_1^b(p) &\equiv& \frac{\pi \sqrt{3}}{128} \, \frac{(p^4\,+\, 4
\,k^2\,+\,3)^2\,(w_p-w_b) (1\,+\,n_b\,+\,+n_p)}{
w_p^3 (w_p+w_b)} {\rm sech}^2 \left[\frac{\pi \,p}{2}\right] \nonumber \\
\Gamma_2^b(p) &\equiv& \frac{\pi \sqrt{3}}{128} \, \frac{(p^4\,+\, 4\,k^2\,
+\,3)^2\,(w_p+w_b) (n_b\,-\,n_p)}{
w_p^3 (w_p-w_b)} {\rm sech}^2 \left[\frac{\pi \,p}{2}\right] \nonumber \\
\Sigma_1(p,k) & \equiv & (w_p\,+\,w_k)\;\Gamma_1(p,k) \; ; \; \Sigma_2(p,k)
=  (w_p\,-\,w_k)\;\Gamma_2(p,k) \nonumber \\
\Sigma_1^b(p) & \equiv & (w_p\,+\,w_b)\;\Gamma_1^b(p) \; ; \; \Sigma_2^b(p)
=  (w_p\,-\,w_b)\;\Gamma_2^b(p) \nonumber \\
w_p^2 & = & p^2\,+\,4 \; ,
\end{eqnarray}
where $\Sigma(\tau)$ and $\Gamma(\tau)$ are defined as in eqn.(\ref{dimless}). 
The functions
$\Sigma(\tau)$ and $\Gamma(\tau)$ where evaluated numerically at different 
temperatures ${\cal T}$, the results are displayed in fig.(\ref{figpsg}). 
The behavior of these functions differ from those in the Sine-Gordon theory
because of the presence of the bound state which is interpreted as an excited
state of the soliton.
As the soliton moves in the dissipative medium, energy is transferred between
the soliton and the bound state resulting in the Rabi-like  oscillations 
displayed in the figure. We notice that the contribution of the bound state 
is of the
same order of magnitude as that of the continuum.

\subsubsection{ Equation of Motion: exact solution vs. Markovian approximation}

The solution to the equation of motion and the comparison to the Markovian
approximation proceeds just as in the the case of the Sine-Gordon model.
The equation of motion is again solved for the case of a constant electric
field (after switching on at $t=0$). The exact and Markovian solutions are 
displayed in figure (\ref{figpfv}).

The new feature of the solution are the oscillations that result from virtual
transitions to the bound state. We interpret these in the following manner:
as the soliton moves it excites the bound state that corresponds to
a soliton distortion, this excitation in turn reacts-back in the dynamics
of the collective coordinate in a retarded manner. 

 While the exact solution  in this model is qualitatively  similar to that
of the Sine-Gordon model, we see however, that quantitatively they are 
different: there
is stronger dynamical dissipation in the $\phi^4$ model as compared to the
Sine Gordon case, due to the strong coupling to the bound state-continuum 
intermediate states. Figure (\ref{figpfv})reveals that whereas the oscillations
arising from the
excitation of the bound state are not very noticeable in the exact solution,
the Markovian approximation is very sensitive to these oscillations and provides a
 misrepresentation of the dynamics. We infer from this analysis that the
memory terms are very important and cannot be neglected.

\subsubsection{ Velocity relaxation and wave function renormalization}

In this model the Laplace transform of the functions $\Gamma(\tau)$ and 
$\Sigma(\tau)$ are given by
\begin{eqnarray}
\tilde{\Sigma}(s) & = & \int_{-\infty}^\infty 
{\rm d} p \, {\rm d} k \left\{ 
{\Sigma_1(p,k)\;s \over{s^2 + (w_p\,+\,w_k)^2} } \,+ \,
{\Sigma_2(p,k)\;s \over{s^2 + (w_p\,-\,w_k)^2} } \right\} \nonumber \\
& & \quad + \quad
\int_{-\infty}^\infty 
{\rm d} p  \left\{ 
{\Sigma_1^b(p)\;s \over{s^2 + (w_p\,+\,w_b)^2} } \,+ \,
{\Sigma_2^b(p)\;s \over{s^2 + (w_p\,-\,w_b)^2} } \right\} \nonumber \\
\tilde{\Sigma}(s) & \equiv & s\, \tilde{\Gamma}(s) \nonumber \\
\tilde{\Gamma}(s) & = & \int_{-\infty}^\infty 
{\rm d} p \, {\rm d} k \left\{ 
{\Gamma_1(p,k)\;(w_p\,+\,w_k) \over{s^2 + (w_p\,+\,w_k)^2} } \,+ \,
{\Gamma_2(p,k)\;(w_p\,-\,w_k) \over{s^2 + (w_p\,-\,w_k)^2} } \right\} \nonumber \\
& & \quad + \quad
\int_{-\infty}^\infty 
{\rm d} p  \left\{ 
{\Gamma_1^b(p)\;(w_p\,+\,w_b) \over{s^2 + (w_p\,+\,w_b)^2} } \,+ \,
{\Gamma_2^b(p)\;(w_p\,-\,w_b) \over{s^2 + (w_p\,-\,w_b)^2} } \right\}. 
\end{eqnarray}

With the quantities $\Sigma^b \; ; \; \Gamma^b$ given above. 
The homogeneous equations of motion given by (\ref{seem})(exact)  and its 
Markovian approximation (\ref{smem}) both with $j=0$ are  solved with the 
kernels $\Sigma(\tau) \; ; \; \Gamma(\tau)$ given above with initial 
condition $v_0=1$. The asymptotic behavior of
the  exact solution will be compared with the prediction $v_{\infty}/v_0
= Z_s$, with
the wave function renormalization $Z_s$ given by eqn.(\ref{wavefc}) but with
the $\tilde{\Gamma}(s=0)$ appropriate to the $\phi^4$ model.

Fig.(\ref{figpv}) shows the numerical solutions of eqn.(\ref{seem}) and 
eqn.(\ref{smem}) with $j=0$  for temperatures ${\cal T}$ = 0, 1.0, 5.0 and
10.0 with initial condition $v_0=1$. Again the Rabi-like oscillations 
associated with
the excitation of the bound state is apparent in the solutions. We have 
checked numerically that asymptotically the velocity tends to a constant value
$v_{\infty}$ but not
exponentially.  Table 2 shows the values of  $v_{\infty}$ and $Z_s$ for  these
temperatures for $m/M\,=\,$0.1 and 0.25 where $v_\infty$ was evaluated at $\tau
= 200$ for the exact solution with $v_0=1$. Within our numerical errors, we can
see that eqn.(\ref{wavefc}) is fulfilled.

\begin{center}
\begin{tabular}{||c||c|c||c|c||} 
\multicolumn{5}{c}{ \textbf{Table (2)} } \\ 
\multicolumn{5}{c}{  } \\ \hline \hline
   & \multicolumn{2}{c||}{ $v_\infty / v_0$ } & \multicolumn{2}{c||}
{$Z_s$ } \\ \hline
   & m/M = 0.1 & m/M = 0.25& m/M = 0.1 & m/M = 0.25 \\  \hline\hline
Zero Temp. &  0.999225 & 0.998064 & 0.999176 & 0.997943 \\ \hline
Temp. 1.0 &  0.961561 & 0.911004 & 0.96376 &  0.914072 \\ \hline
Temp. 5.0 &  0.784934 & 0.593058 &  0.787734 & 0.597494 \\ \hline
Temp. 10.0 & 0.642698 & 0.417231 & 0.646577 & 0.422562 \\ \hline \hline 
\end{tabular}
\end{center}

\subsubsection{ Kernels for the semiclassical Langevin equation }

>From the definition of the kernels $K_1(t-t^\prime)$ and $K(t-t^\prime)$,
eqns.(\ref{ker1}) and (\ref{ker}), and eqn.(\ref{TaTsp4}), these kernels
can be written in terms of the dimensionless quantities given by 
eqn.(\ref{change}) as
\begin{equation}
{\cal K}_1(\tau-\tau^\prime)\,=\, -2\,i\,\Sigma(\tau-\tau^\prime),
\end{equation}
where $\Sigma(\tau-\tau^\prime)$ is given by eqn.(\ref{fisig}) and 
$K(t) = m^4 {\cal K}(\tau)$ with 
\begin{eqnarray}
{\cal K}(\tau)&=& \int_{-\infty}^\infty {\rm d} p \biggl\{ \; C_1^b(p)\,
\cos\left[(w_p + w_{b})(\tau ) \right]\;+\;C_2^b(p)\,\cos\left[(w_p - w_{b})
(\tau ) \right] \nonumber \\
& & + \quad 
\int_{-\infty}^\infty{\rm d} k \left\{C_1(p,k)\,\cos\left[(w_p + w_{k})(\tau)
\right]\;+\;C_2(p,k)\,\cos\left[(w_p - w_{k})(\tau) \right] \right\}
\biggr\} \nonumber \\
\end{eqnarray}
with the dimensionless matrix elements 
\begin{eqnarray}
C_1(p,k) &\equiv& {18 \over{4^5}}\;{( 1 + n_p + n_k + n_p\,n_k) (p^2 - k^2)^4
(p^2 + k^2 + 4)^2 
\over{w_p^3\,w_k^3\,(p^2+1)\,(k^2+1)\sinh^2\left[ {\pi \over 2} 
(p + k)\right]}} \nonumber \\
C_2(p,k) &\equiv& {9\over{4^4}}\;{n_k(1 + n_p) (p^2 - k^2)^4 (p^2 + k^2 + 4)^2 
\over{w_p^3\,w_k^3\,(p^2+1)\,(k^2+1)\sinh^2\left[ {\pi \over 2} (p + k)\right]}}
\nonumber \\
C_1^b(p) &\equiv& \frac{\pi \sqrt{3}}{4^4} \, \frac{(p^4\,+\, 4\,k^2\,+\,
3)^2\,(k^2+1) (1 + n_b + n_p + n_b n_p)}{
w_p^3  {\rm cosh}^2 \left[\frac{\pi \,p}{2}\right]} \nonumber \\
C_2^b(p) &\equiv& \frac{2 \pi \sqrt{3}}{4^4} \, \frac{(p^4\,+\, 4\,k^2\,+\,
3)^2\,(k^2+1) n_k(1+\,n_b)}{w_p^3 {\rm cosh}^2 \left[\frac{\pi \,p}{2}\right]}. 
\end{eqnarray}

Fig.(\ref{figpcr}) shows ${\cal K}(\tau)$ vs. $\tau$  for temperatures 
${\cal T}=0,1,5,10$. Again the oscillations
are a consequence of the bound state contribution, and as in the Sine-Gordon
case we find that despite the fact that in the high temperature limit the 
kernel becomes very localized in time, the total integral 
$\int_{-\infty}^{\infty}d\tau {\cal K}(\tau) =0$ preventing
a representation of the noise-noise correlation function as a delta 
function in time even in the high temperature limit, which for example 
for {\em trans}-polyacetylene is beyond the experimentally relevant scales.
The ``color'' in the noise-noise
correlation function is enhanced by the strong coupling to the continuum via
the bound state which is also responsible for the strong oscillatory behavior
of the real-time correlation function. 

The high temperature limit,  $T>>m$ also implies a breakdown of the
adiabatic (perturbative) expansion. In this limit the relevant scale in the
kernels is $T$ and the rescaling of variables in (\ref{change}) should be in
terms of $T$ rather than $m$. This implies that the expansion is now in terms
of the ratio $T/M$ which for high temperatures will imply strong coupling for
the models under consideration when the parameters are fixed to make contact 
with the materials of interest.  This is common to both the $\phi^4$ and
Sine-Gordon models, leading us to conclude that the classical limit requires
a non-perturbative approach, which is clearly beyond
the realm of this work.

\section{ Conclusions and further questions}

We have studied the non-equilibrium dynamics of solitons by obtaining the
real time equations of motion for the expectation value of the collective
coordinate and also the quantum Langevin equation to lowest order in the adiabatic
expansion. These allowed us to obtain the frequency dependent soliton conductivity
in this expansion. 

 The Hamiltonian for a $\phi^4$ field theory was studied as a model
for conducting polymers and the Sine-Gordon model was used to describe the phase
soliton dynamics for charge density wave systems. In both cases parameters 
were chosen to describe the experimental realizations of these systems. 

To lowest order in the adiabatic coupling we found that the real-time
equation of motion involves a non-Markovian self-energy kernel and that the
static friction coefficient vanishes. However, there is dynamical friction
which is a result of the memory effects in the
self-energy and is associated with two-phonon processes. We studied the 
Markovian approximation and shown numerically that this approximation is 
unreliable in the relevant range of temperatures. 

The quantum Langevin equation was obtained by computing the influence functional
obtained by tracing out the phonon (or phason) degrees of freedom to the same 
order in the adiabatic expansion. We found that the dissipative kernel and the noise 
correlation function obey a generalized form of fluctuation-dissipation relation
but that a Markovian limit is not available, the noise is gaussian, additive
but colored. We have studied the high temperature limit to
establish whether the classical limit emerges, and found that
to first adiabatic order  and for the experimentally relevant range of 
temperatures, the
classical limit of these kernels is not achieved. We pointed
out that formally the high temperature
limit leads to a breakdown of the adiabatic expansion and requires
a non-perturbative treatment. 

There are several possible avenues to pursue: a higher order calculation for
example as carried out by Ogata et. al.\cite{ogata} but implemented in real-time
to obtain the  non-equilibrium evolution of solitons and the associated quantum
Langevin equation with a detailed study of the classical limit. 

In the case of {\em trans}-polyacetylene the adiabatic ratio is not so small and
a perturbative (adiabatic) expansion could be deemed suspect and certainly 
untrustworthy in the high temperature regime. A possible avenue to pursue in this
case would be a variational calculation with a few variational parameters, one
of them would be the collective coordinate and others related to the soliton
distortion. Such  treatment would also be valuable to study the situation of 
large soliton velocities which necessarily imply a ``Lorentz contraction'' of the
soliton profile. In this case a more realistic model to study would be the 
continuum model of Takayama, Liu and Maki\cite{tlm} that incorporates the
electronic degrees of freedom, which we expect would add quantitative changes to 
the dissipative contributions. 
Furthermore, in order to establish a close connection with experiments, the
effects of impurities and other pinning potentials must be understood. 

Work on some of these aspects is in progress.

\noindent{\bf Acknowledgements}
The authors would like to thank  D. Jasnow, H. deVega, J. Levy, R. Willey and D. 
Campbell for helpful discussions and comments. They also thank N.S.F. for partial
support through grant awards: PHY-9605186, INT-9512798 and the Pittsburgh 
Supercomputing Center for grant: PHY-PHY950011P. S. A. thanks King Fahad 
University of Petroleum and Minerals (Saudi Arabia) for
financial support. F. I. T. thanks the Dept. of Physics, Univ. of Pittsburgh
for hospitality and CNPq and FAPEMIG for financial support.   

%\section*{ Appendices}

%\renewcommand{\theequation}{\thesection.\arabic{equation}}
%\setcounter{equation}{0}

%\setcounter{equation}{0}

%\setcounter{equation}{0}
\appendix

\section{Real-Time Phonon Correlation Functions \label{B}}

In this appendix, we will calculate the Green's functions which are defined 
in eqn.(\ref{greens}) in terms of the vertex given by
eqn. (\ref{Faa}). 

Applying Wick's theorem and eqn.(\ref{propaa}), it is a matter of
 straightforward algebra to find the following results:
\begin{eqnarray}
G^{++}(t,t^\prime) & = &  -2 \,\sum_{p,k\not=0} \Biggl\{ T_{pk}^{(S)} \,
T_{-p-k}^{(S)} \biggl[ \,e^{-i (\omega_p + \omega_k)(t - t^\prime) } \Bigl( 
n_p n_k + \theta(t - t^\prime)( 1 +  n_p + n_k ) \Bigr) \nonumber \\
& & \hspace{1.2in} + \quad e^{+ i (\omega_p + \omega_k)(t - t^\prime) } 
\Bigl( n_p n_k + \theta(t^\prime - t)( 1 +  n_p + n_k ) \Bigr) \biggr] 
\nonumber \\
& & \quad \;  + \;\; 2\, T_{pk}^{(A)} \,
T_{-p-k}^{(A)} \biggl[e^{- i (\omega_p - \omega_k)(t - t^\prime) } 
\Bigl( n_p n_k + n_p\, \theta(t^\prime - t) +  n_k \, \theta(t - t^\prime)
\Bigr) \biggr] \Biggr\}
\nonumber \\
& & \nonumber \\
& & \nonumber \\
G^{--}(t,t^\prime) & = &  -2 \,\sum_{p,k\not=0} \Biggl\{ T_{pk}^{(S)} \,
T_{-p-k}^{(S)} \biggl[ \,e^{-i (\omega_p + \omega_k)(t - t^\prime) } 
\Bigl( n_p n_k + \theta(t^\prime - t)( 1 +  n_p + n_k ) \Bigr) \nonumber \\
& & \hspace{1.2in} + \quad e^{+ i (\omega_p + \omega_k)(t - t^\prime) } 
\Bigl( n_p n_k + \theta(t - t^\prime)( 1 +  n_p + n_k ) \Bigr) \biggr] 
\nonumber \\
& & \quad \;  + \;\; 2\, T_{pk}^{(A)} \,
T_{-p-k}^{(A)} \biggl[e^{- i (\omega_p - \omega_k)(t - t^\prime) } \Bigl( 
n_p n_k + n_k\, \theta(t^\prime - t) +  n_p \, \theta(t - t^\prime) 
\Bigr) \biggr] \Biggr\}
\nonumber \\
& & \nonumber \\
& & \nonumber \\
G^{+-}(t,t^\prime) & = &  2 \,\sum_{p,k\not=0} \Biggl\{ T_{pk}^{(S)} \,
T_{-p-k}^{(S)} \biggl[ \,e^{-i (\omega_p + \omega_k)(t - t^\prime) } \; 
n_p n_k \, + \, e^{+ i (\omega_p + \omega_k)(t - t^\prime) }\Bigl( n_p n_k 
+ n_p + n_k + 1) \Bigr) \biggr]\nonumber \\
& & \quad \;  + \;\; 2\, T_{pk}^{(A)} \,
T_{-p-k}^{(A)} \biggl[e^{- i (\omega_p - \omega_k)(t - t^\prime) } \; n_p 
\Bigl(1 +  n_k \Bigr) \biggr] \Biggr\} \nonumber \\
& & \nonumber \\
& & \nonumber \\
G^{-+}(t,t^\prime) & = & 2 \,\sum_{p,k\not=0} \Biggl\{ T_{pk}^{(S)} \,
T_{-p-k}^{(S)} \biggl[ \,e^{-i (\omega_p + \omega_k)(t - t^\prime) } \Bigl( 
n_p n_k + n_p + n_k + 1) \, + \, e^{+ i (\omega_p + \omega_k)(t - t^\prime) }
\; n_p n_k  \biggr]\nonumber \\
& & \quad \;  + \;\; 2\, T_{pk}^{(A)} \,
T_{-p-k}^{(A)} \biggl[e^{- i (\omega_p - \omega_k)(t - t^\prime) } \; n_k 
\Bigl(1 +  n_p \Bigr) \biggr] \Biggr\}.  
\label{GREENS}
\end{eqnarray}

These Green's functions satisfy the following relation
\begin{equation}
G^{++} \, + \, G^{--} \, + \,G^{+-} \, + \,G^{-+} \, = \,0
\end{equation}
which is a consequence of unitary time evolution\cite{b6}.

Furthermore, using the antisymmetry property of the matrix elements
$T^{(A)}_{pk}$ one finds that 
\begin{equation}
G^{+-}(t,t') = (G^{-+}(t,t'))^* \label{conju}
\end{equation}
The Green's functions $G^{++}(t,t')\; ; \; G^{--}(t,t')$ can be
written in terms of $G^{+-}(t,t')$ and its complex conjugate, therefore
we see that there is only one independent Green's functions (and its
complex conjugate). 

\section{Calculating $K_1(t-t^\prime)$ and $K(t-t^\prime)$ \label{C} }

Performing the coordinate transformation in eqn.(\ref{wigner}), the 
influence-functional becomes
\begin{eqnarray}
{\cal F}[\dot{x},\dot{R}] & = & exp\left\{-\, \frac{1}{2} \int dt\, dt^\prime \right. \biggl[ \frac{\dot{R}(t) \,\dot{R}(t^\prime)}{4} \biggl(  G^{++}(t,t^\prime) \,+\, G^{--}(t,t^\prime) \,-\, G^{+-}(t,t^\prime)\,-\, G^{-+}(t,t^\prime) \biggr) \nonumber \\ 
& + &  \biggl\{
\frac{1}{2} \dot{R}(t) \,\dot{x}(t^\prime) \biggl(  G^{++}(t,t^\prime) \,-\,
G^{--}(t,t^\prime) \,+\, G^{+-}(t,t^\prime)\,-\, G^{-+}(t,t^\prime) \biggr)
\nonumber \\
& + &  \left. \;
\frac{1}{2} \dot{x}(t) \,\dot{R}(t^\prime) \biggl(  G^{++}(t,t^\prime) \,-\,
G^{--}(t,t^\prime) \,-\, G^{+-}(t,t^\prime)\,+\, G^{-+}(t,t^\prime) \biggr)
\biggr\} \biggr] \right\}.
\end{eqnarray}

Integrating the linear term in $\dot{R}$ by parts once and the quadratic term 
twice, the influence-functional can be cast in the following form
\begin{equation}
{\cal F}[\dot{x},\dot{R}]  =  exp\left\{\, \frac{1}{2} \int dt\, dt^\prime \Bigl[
R(t) \, K_1(t-t^\prime) \,\dot{x}(t^\prime) \,-\, R(t) \,K(t-t^\prime)\, \dot{R}
(t^\prime) \Bigr] \right\},
\end{equation}
where
\begin{eqnarray}
K_1(t-t^\prime) & = & \frac{1}{2} \frac{\partial}{\partial \, t} \biggl[\Bigl(
G^{++}(t,t^\prime) \,-\, G^{--}(t,t^\prime) \,+\, G^{+-}(t,t^\prime)\,-\, G^{-+}
(t,t^\prime) \Bigr) \nonumber \\
& & + \quad \;\; \Bigl(  G^{++}(t^\prime,t) \,-\, G^{--}(t^\prime,t) \,-\, G^{+-}
(t^\prime,t)\,+\, G^{-+}(t^\prime,t) \Bigr) \biggr] \\
K(t-t^\prime) & = & \frac{1}{4} \frac{\partial^2}{\partial \, t^2} \biggl[ G^{++}
(t,t^\prime) \,+\, G^{--}(t,t^\prime) \,-\, G^{+-}(t,t^\prime)\,-\, G^{-+}
(t,t^\prime) \biggr].
\end{eqnarray}

The generalized fluctuation-dissipation relation is obtained by writing
the two kernels above in terms of $G^{\pm}(t,t')$, the only independent
Green's function. 

Substituting the values of the Green's functions from eqn.(\ref{GREENS}) in the
above equations, one obtains the expressions for $K_1(t-t^\prime)$ and 
$K(t-t^\prime)$ in eqns.(\ref{ker1}) and (\ref{ker}).

In the case that there are bound states other than the zero mode, such as
the case of $\phi^4$ the sum in eqn.(\ref{Faa}) runs over all bound and 
scattering states, i.e.
\begin{eqnarray}
F[a^\dagger, a] &=&  \frac{1}{2\,i} \int dp\,dk \sqrt{ {\omega_p 
\over{\omega_k}}} D_{pk} \left[ a_k a_p - a_{-k}^\dagger 
a_{-p}^\dagger + a_{-k}^\dagger a_p - a_{-p}^\dagger a_k \right] \nonumber \\
& & + \;\frac{1}{2\,i} \sum_b \int dk \sqrt{ {\omega_b 
\over{\omega_k}}} D_{bk} \left[ a_k a_b - a_{-k}^\dagger 
a_{b}^\dagger + a_{-k}^\dagger a_b - a_{b}^\dagger a_k \right] \nonumber \\
& & + \; \frac{1}{2\,i} \sum_b \int dk \sqrt{ {\omega_k 
\over{\omega_b}}} D_{kb} \left[ a_b a_k - a_{b}^\dagger 
a_{-k}^\dagger + a_{b}^\dagger a_k - a_{-k}^\dagger a_b \right] \nonumber \\
& & + \; \frac{1}{2\,i} \sum_{a,b} \sqrt{ {\omega_a 
\over{\omega_b}}} D_{ab} \left[ a_b a_a - a_{b}^\dagger 
a_{a}^\dagger + a_{b}^\dagger a_a - a_{a}^\dagger a_b \right], \label{lll}
\end{eqnarray}
where the indices $a$ and $b$ stand for summation over discrete bound states 
and $p$ and $k$ stand for summation over continuum scattering states. 
The models which we considered in this paper have at most one bound state, 
that is the case in the $\phi^4$ theory. In this case, the last term will not
contribute since $D_{bb}$ vanishes. Thus for only one bound state, 
eqn.(\ref{lll}) can be written as
\begin{eqnarray}
F[a^\dagger, a] &=& \int dp\,dk\, \left[T^{(S)}_{pk}\,\left(a_p\,a_k \,-\,
a^\dagger_{-p}\,
a^\dagger_{-k}\right)\,+\,T^{(A)}_{pk}\,\left(a^\dagger_{-p}\,a_k \,-\,
a^\dagger_{-k}\,
a_{p}\right) \right]\nonumber \\
& & + \;
\int dk\, \left[T^{(S)}_{bk}\,\left(a_k\,a_b \,-\,a^\dagger_{-k}\,
a^\dagger_{b}\right)\,+\,T^{(A)}_{bk}\,\left(a^\dagger_{-k}\,a_b \,-\,
a^\dagger_{b}\,
a_{k}\right) \right], \label{foobd}
\end{eqnarray}
where the matrices $T^{(S)}_{pk}$ and $T^{(A)}_{pk}$ for scattering states
are given by eqn.(\ref{scm})  
and if one of the states is a bound state, then
\begin{eqnarray}
T_{bk}^{(S)} &=& {1\over{2 i}} \left[\sqrt{{\omega_b 
\over{\omega_k}}} - \sqrt{{\omega_k\over{\omega_b}}}\, 
\right] D_{bk} \nonumber \\
T_{bk}^{(A)} &=& {1\over{2 i}} \left[\sqrt{{\omega_b
\over{\omega_k}}} + \sqrt{{\omega_k\over{\omega_b}}}\, 
\right] D_{bk} \,.
\label{Tsa}
\end{eqnarray}

In the Sine-Gordon theory, the last two terms in eqn.(\ref{foobd}) do not 
contribute since in this theory there are no bound states other than the 
zero mode and the Green's functions are given by eqn.(\ref{GREENS}) but 
with integration over $p$ and $k$ instead of the summation.

In the $\phi^4$ case, to lowest adiabatic order the contributions from the 
bound and scattering states decouple. This implies that  the Green's 
functions will have a contribution from the bound state which is given 
by the same expression as that of the scattering states, with 
$p \rightarrow b$, but multiplied by a factor of $1/2$ since the bound 
state wave function is chosen to be real.

%%%%%%%%%%%%%%%%%%bibliography here
%\input{solibib}

%%%%%%%%%%%%%%%%%end of bibliography

\setlength{\textheight}{8.4in}

%%%%%%%%%%%%%%%FIGURES %%%%%%%%%%%%%%%%%%%%%
%%%%%FIGURE 1%%%%%%%%%
\begin{center}
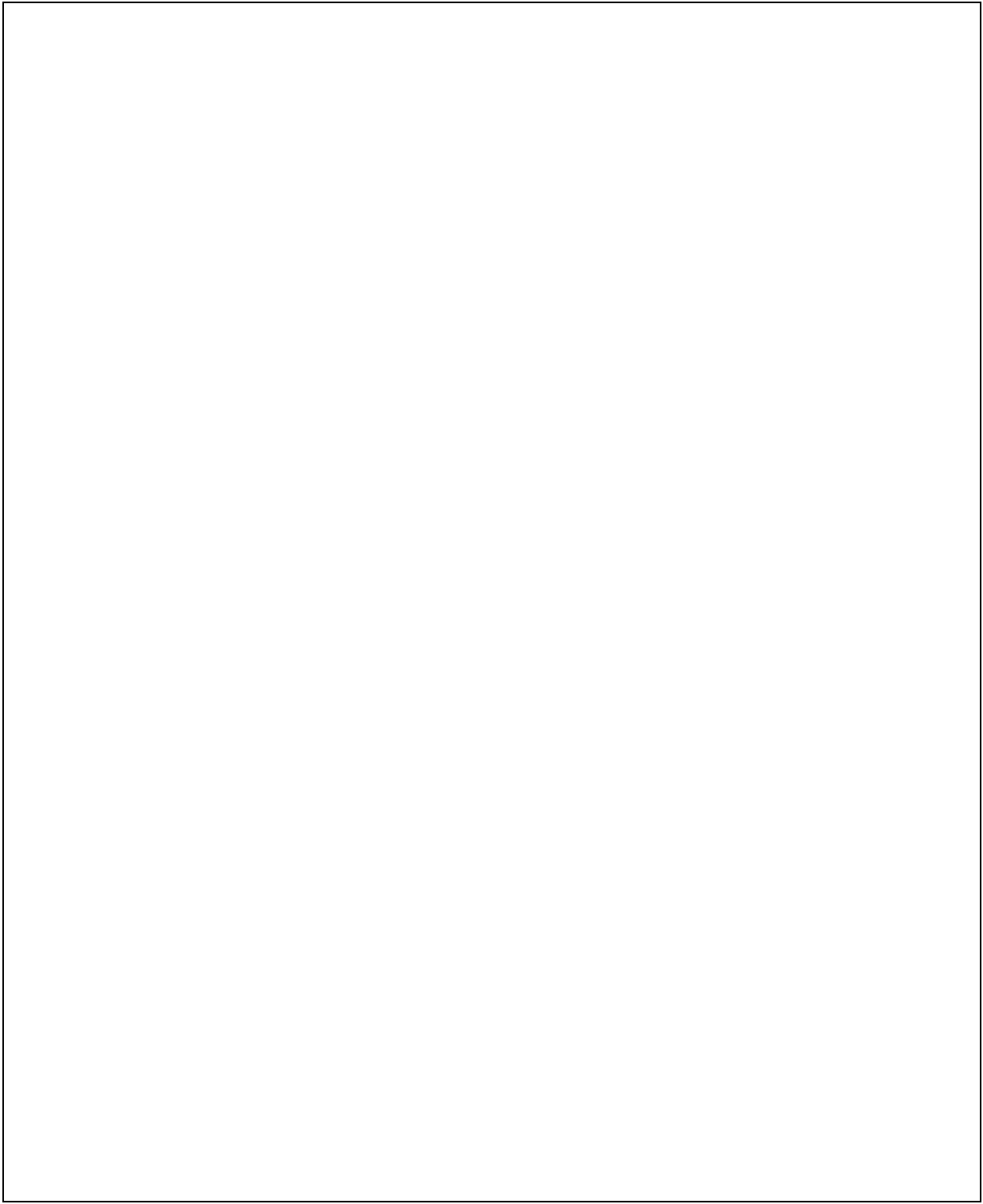
\begin{figure}[t]
\fbox{
\setlength{\unitlength}{1mm}
\begin{picture}(160,200)
\end{picture}}
\caption{The functions $\Gamma(\tau)$ and $\Sigma(\tau)$ for temperatures 
${\cal T}=$ 0, 1.0, 5.0 and 10.0 for Sine-Gordon theory. \label{figssg}}
\end{figure}
\end{center}
%%%%%%END FIGURE 1 %%%%%%%

%%%%% BEGINNING FIGURE 2%%%%%%%%%%%%%%%%%%%%%%

\begin{center}
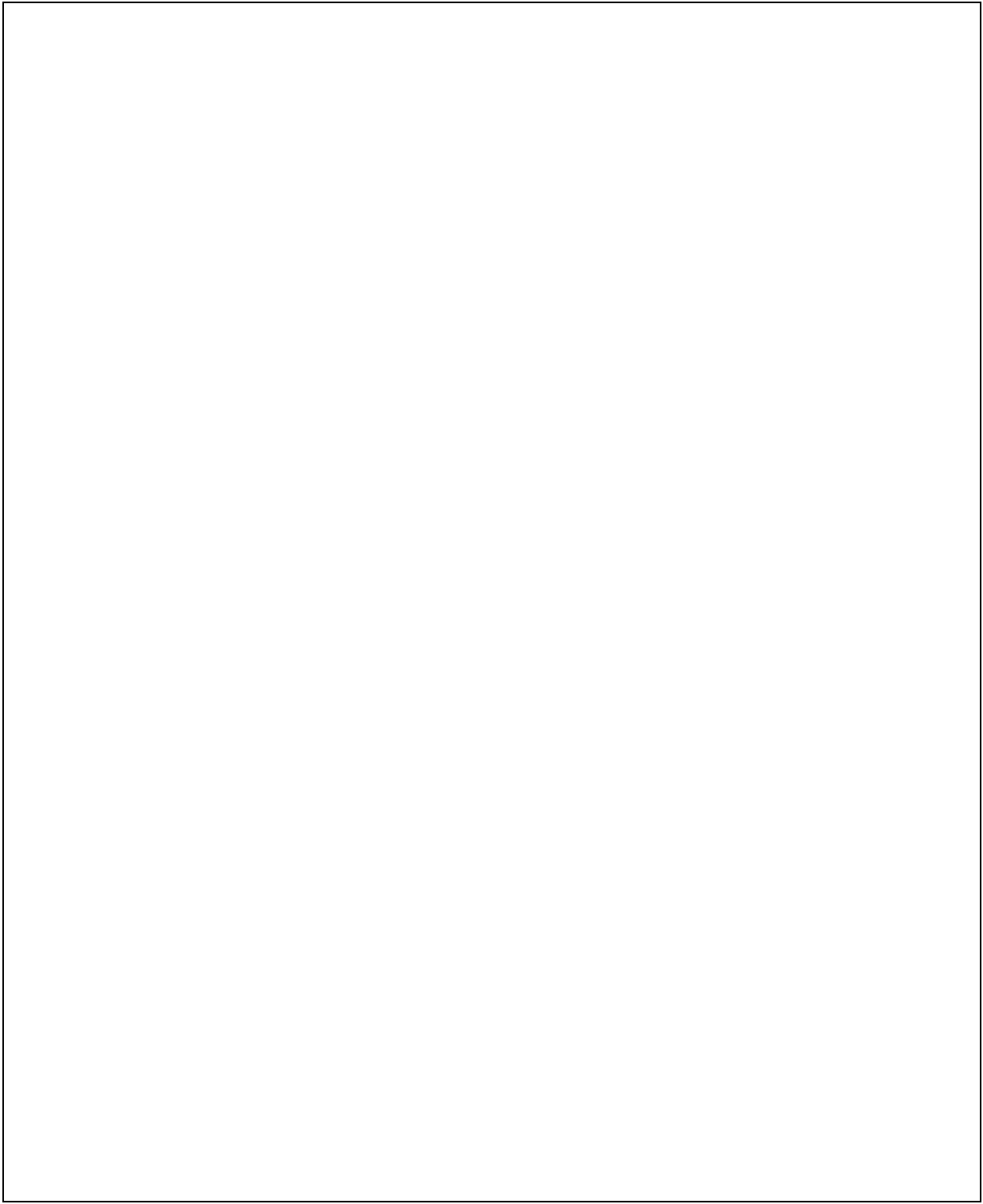
\begin{figure}[t]
\fbox{
\setlength{\unitlength}{1mm}
\begin{picture}(160,200)
\end{picture}}
\caption{Numerical evaluation of the velocity of the soliton in the presence 
of a constant electric field for temperatures ${\cal T}=$ 0, 1.0, 5.0 
and 10.0 in Sine-Gordon theory. \label{figsfv}}
\end{figure}
\end{center}

%%%%%END FIGURE 2 %%%%%%%%%%%%%%%%%%

%%%%%BEGIN FIGURE 3%%%%%%%%%%%

\begin{center}
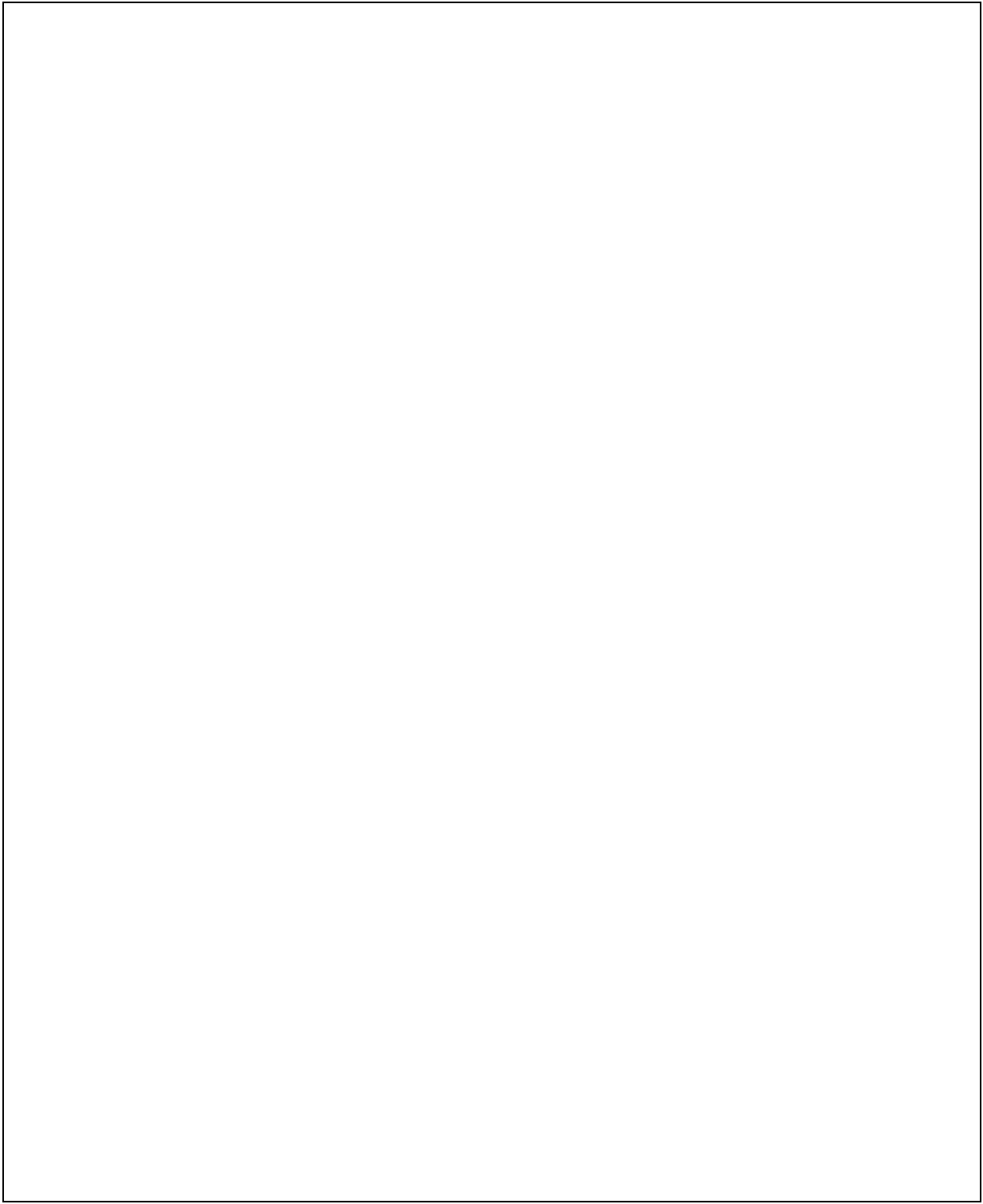
\begin{figure}[t]
\fbox{
\setlength{\unitlength}{1mm}
\begin{picture}(160,200)
\end{picture}}
\caption{ Numerical evaluation of the velocity of the soliton for $j=0 \; ; 
\; v_0=1$ for temperatures ${\cal T}=$ 0, 1.0, 5.0 and 10.0 in Sine-Gordon 
theory. \label{figsv}}
\end{figure}
\end{center}

%%%%%END FIGURE 3 %%%%%%%%%%%%

%%%%%BEGIN FIGURE 4 %%%%%%%%%%%

\begin{center}
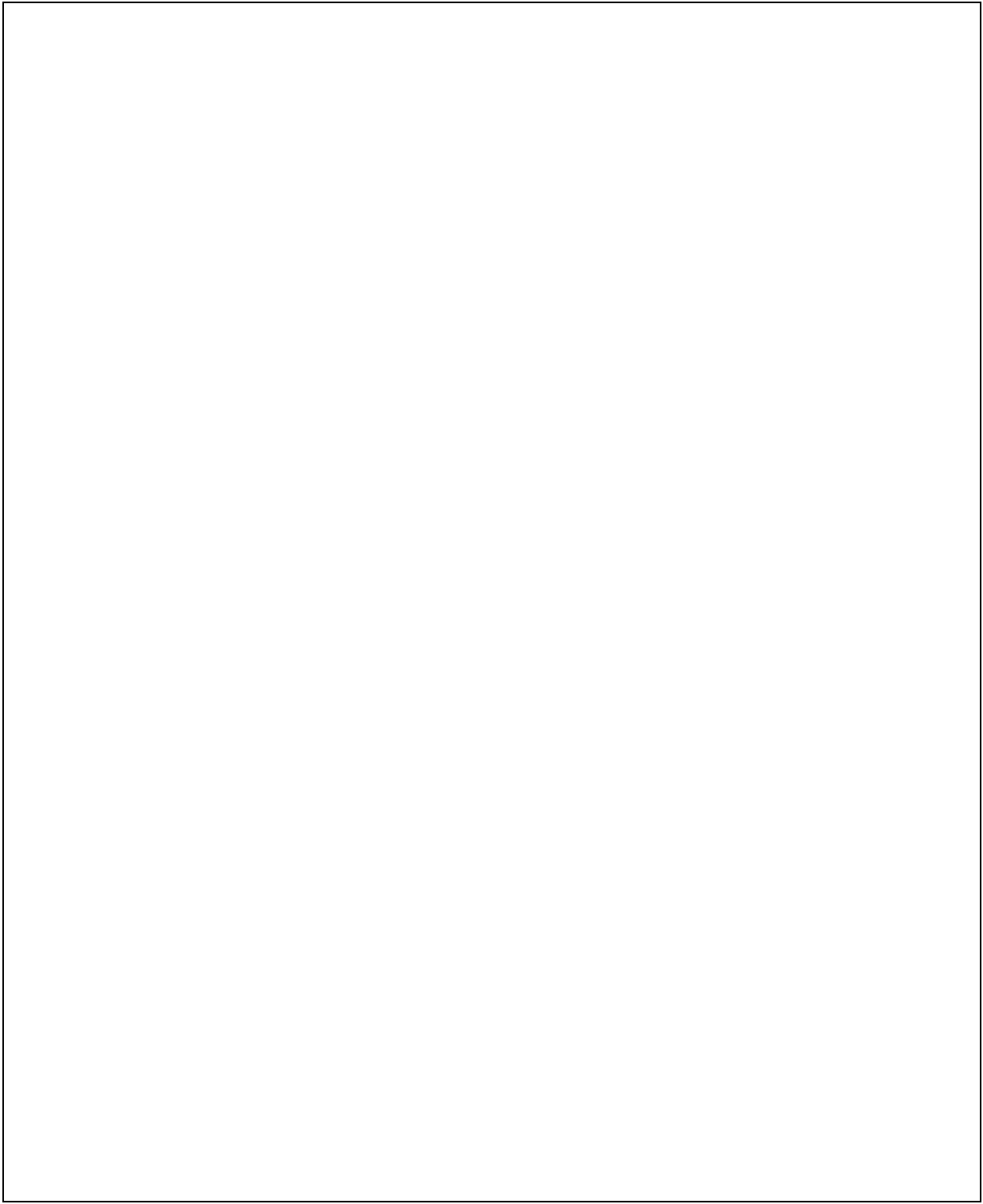
\begin{figure}
\fbox{
\setlength{\unitlength}{1mm}
\begin{picture}(160,200)
\end{picture}}
\caption{ The correlation function ${\cal K}(\tau)$ for temperatures 
${\cal T}=$ 0, 1.0, 5.0 and 10.0 in the Sine-Gordon theory. \label{figscr}}
\end{figure}
\end{center}

%%%%END FIGURE 4 %%%%%%%%%%%%%%%%%%%%

%%%%%%%BEGIN FIGURE 5 %%%%%%%%%

\begin{center}
\noindent 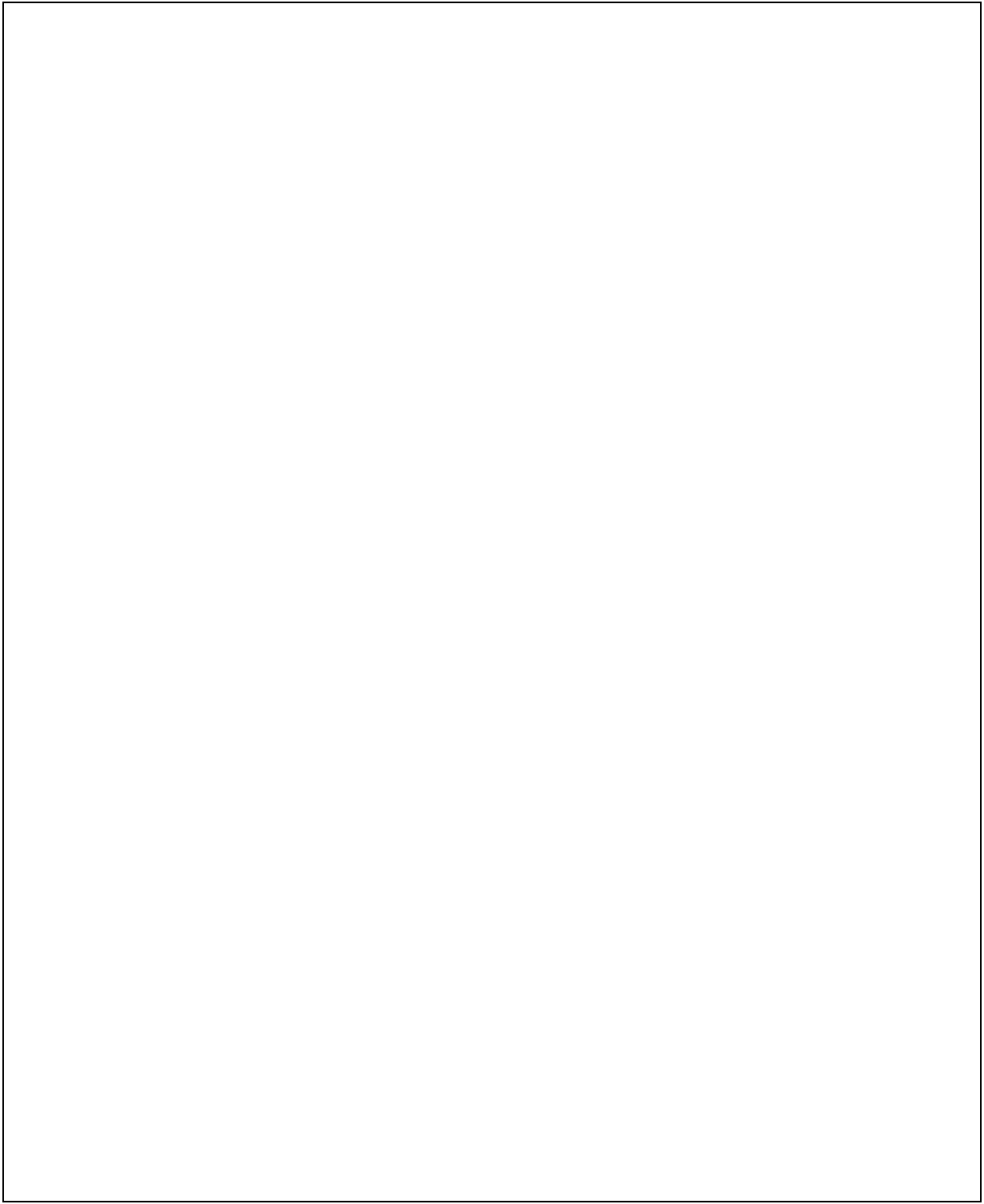
\begin{figure}
\fbox{
\setlength{\unitlength}{1mm}
\begin{picture}(160,200)
\end{picture}}
\caption{The functions $\Gamma(\tau)$ and $\Sigma(\tau)$ for temperatures 
${\cal T}=$ 0, 1.0, 5.0 and 10.0 in the $\phi^4$-theory. Contributions from
bound and scattering states are displayed separately.  \label{figpsg}}
\end{figure}
\end{center}

%%%%%%END FIGURE 5 %%%%%%%%%

%%%%%%%BEGINNING FIGURE 6 %%%%%%%%

\noindent\begin{figure}[t]
\fbox{
\setlength{\unitlength}{1mm}
\begin{picture}(160,200)
\end{picture}}
\caption{ Numerical evaluation of the velocity of the soliton in the presence
of a constant electric field for temperatures ${\cal T}=$ 0, 1.0, 5.0 
and 10.0 in $\phi^4$ theory. \label{figpfv}}
\end{figure}
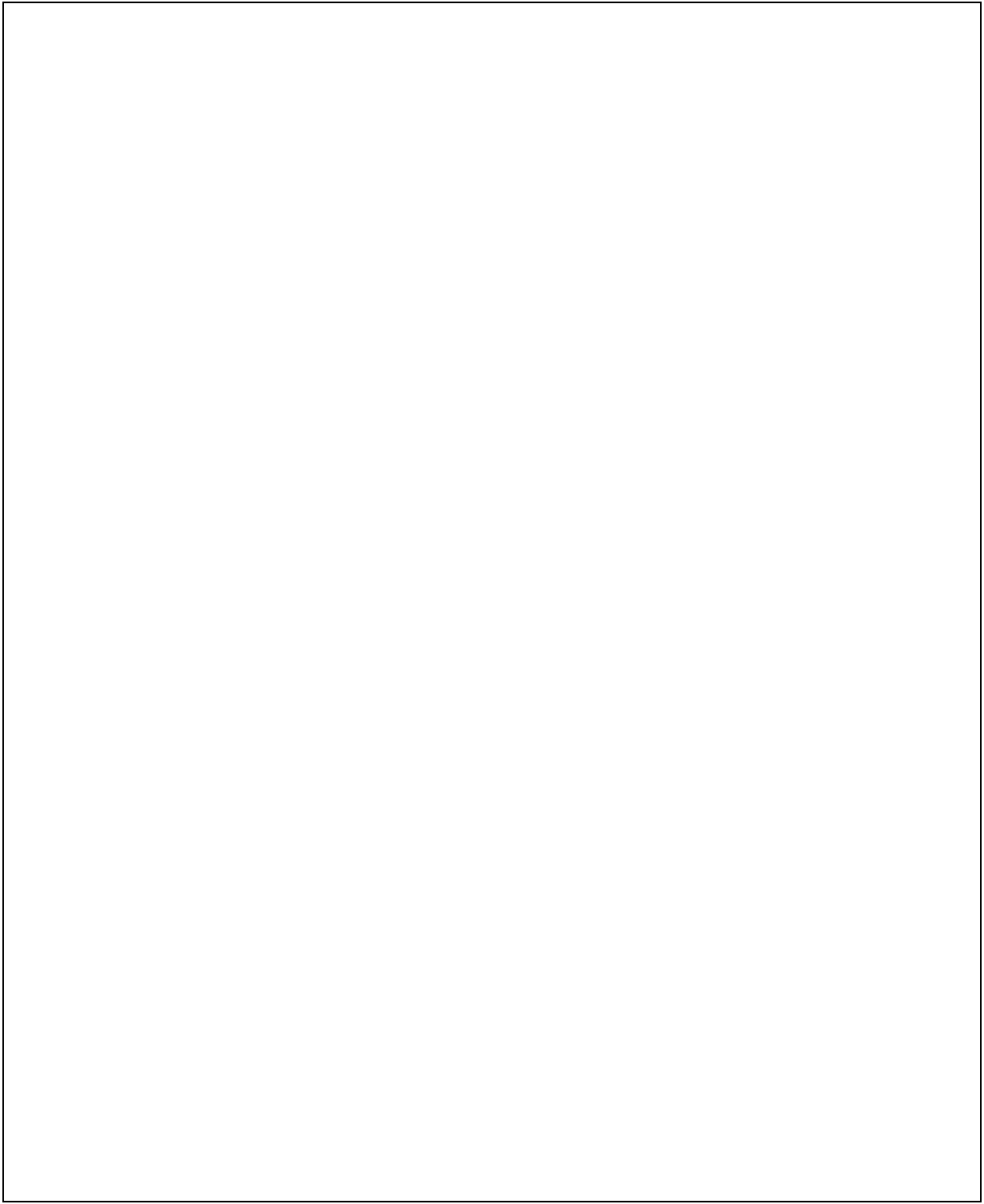

%%%%%END FIGURE 6 %%%%%%%%%%%%%%%%%%%

%%%%%%%%%BEGIN FIGURE 7 %%%%%%%%%

\noindent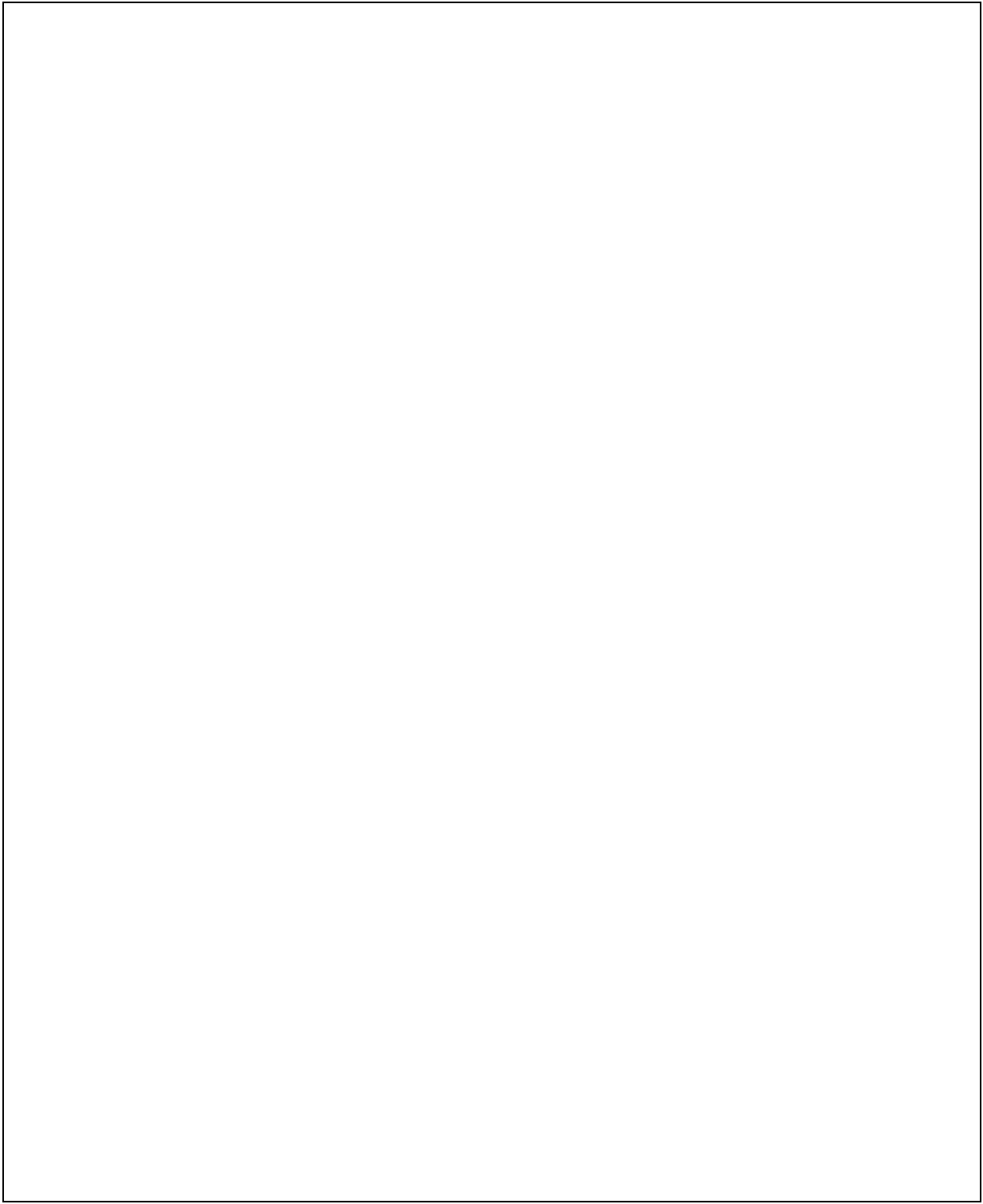
\begin{figure}[t]
\fbox{
\setlength{\unitlength}{1mm}
\begin{picture}(160,200)
\end{picture}}
\caption{Numerical evaluation of the velocity of the soliton for $j=0\; ; 
\; v_0=1$ for temperatures ${\cal T}=$ 0, 1.0, 5.0 and 10.0 in $\phi^4$ 
theory.\label{figpv}}
\end{figure}

%%%%%%%%%%%END FIGURE 7 %%%%%%%%%%%%%

%%%%%%%%%BEGIN FIGURE 8 %%%%%%%%%%%%%%%

\begin{center}
\begin{figure}[t]
\fbox{
\setlength{\unitlength}{1mm}
\begin{picture}(160,200)
\end{picture}}
\caption{The correlation function ${\cal K}(\tau)$ for temperatures 
${\cal T}=$ 0, 1.0, 5.0 and 10.0 in the $\phi^4$ theory. \label{figpcr}}
\end{figure}
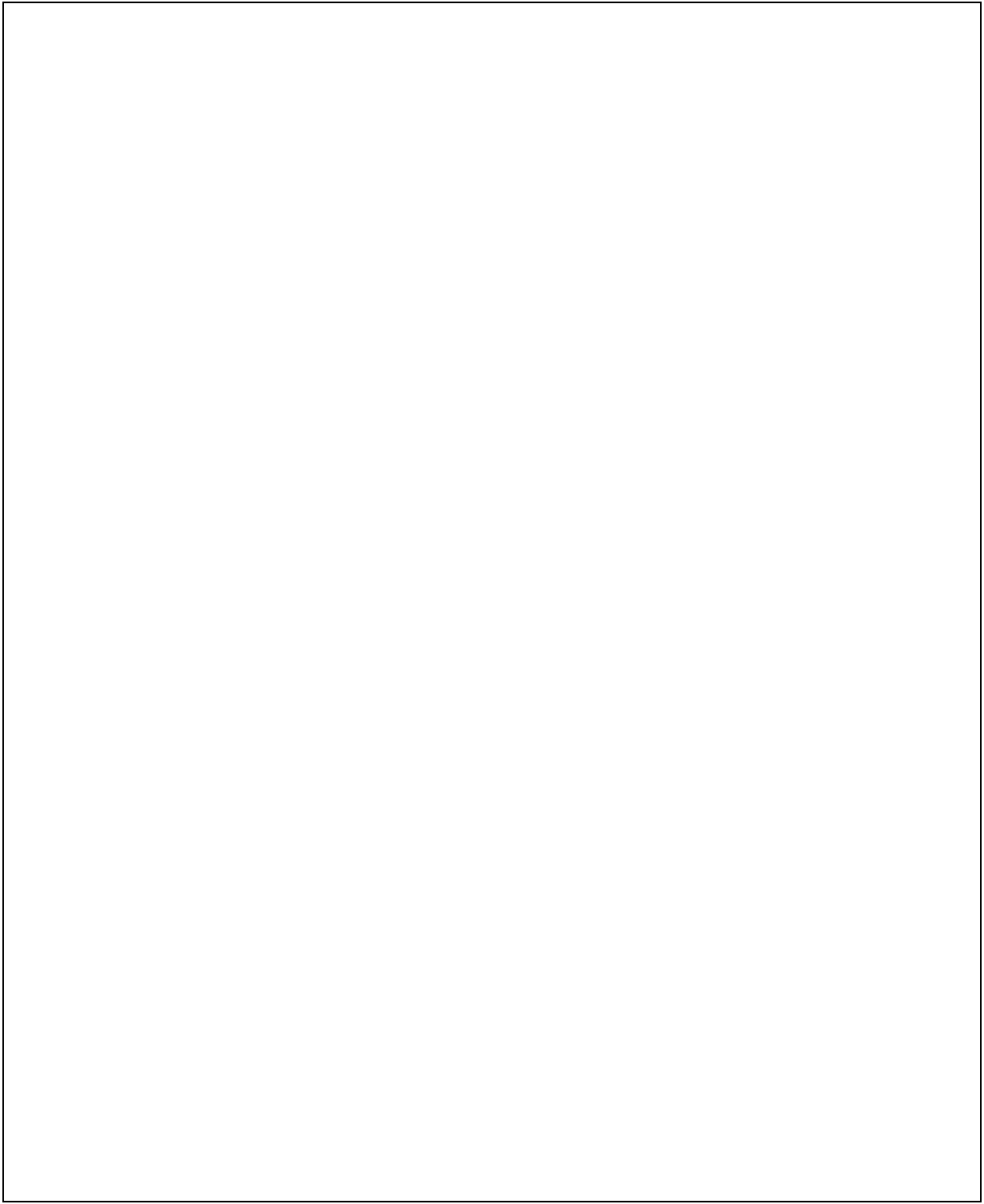
\end{center}

%%%%%%%%%%%%%END FIGURE 8 %%%%%%%%%%%%%%%%%%%%%%%%%

%%%%%%%%%%%%END OF FIGURES%%%%%%%%%%%%%%%%%%%%%

\begin{thebibliography}{99}

\bibitem{krum} J. Krumhansl and J. R. Schrieffer, Phys. Rev. B11, 3535 (1975).
\bibitem{su} W. P. Su, J. R. Schrieffer and A. J. Heeger, Phys. Rev. Lett. 42, 
1698 (1979); Phys. Rev. B22, 2099 (1980). 
\bibitem{schriefrev} A. J. Heeger, S. Kivelson, J. R. Schrieffer and W. P. Su,
Rev. Mod. Phys. 60, 781 (1988).
\bibitem{yu} Yu Lu, ``Solitons and Polarons in Conducting Polymers'', 
(World Scientific, Singapore, 1988)
\bibitem{d3} G. Gruner, {Density Waves in Solids}, Addison-Wesley (1994),
and references therein.
\bibitem{grunerev} G. Gruner, Rev. Mod. Phys. 60, 1129 (1988), and references therein.
\bibitem{rice1} P. A. Lee, T. M. Rice, P. W. Anderson, Solid State Comm. 14, 
703 (1974)
\bibitem{rice2} M. J. Rice, A. R. Bishop, J. A. Krumhansl, S. E. Trullinger,
Phys. Rev. Lett, 36, 432 (1976); M. J. Rice in ``Solitons and Condensed Matter
Physics'', (Eds. by A. R. Bishop and T. Schneider, Springer-Verlag, Berlin, 1978). 
\bibitem{horo} B. Horowitz in ``Solitons'', (Eds. S. Trullinger et. al. North
Holland, Amsterdam, 1986). 
\bibitem{rev} See for example the summary by Y. Wada, Prog. Theor. Phys. Supp.
113, 1 (1993). 
\bibitem{several} T. R. Koehler, A. R. Bishop, J. A. Krumhansl and J. R. 
Schrieffer, Solid State Comm. 17, 1515 (1975). 
\bibitem{wada} Y. Wada and J. R. Schrieffer, Phys. Rev. B18,3897 (1978).
\bibitem{maki} K. Maki, Phys. Rev. B26, 2181; 2187;4539 (1982); Mol. Cryst. Liq.
Cryst. 77, 277 (1981).
\bibitem{tlm} H. Takayama, Y. R. Lin-Liu and K. Maki, Phys. Rev. B21, 2388, (1980).
\bibitem{ogata} M. Ogata and Y. Wada, Prog. Theor. Phys. Supp. 94, 115
(1988); M. Ogata, A. Terai and Y. Wada, Jour. of the Phys. Soc. of Japan,
55, 2305 (1986); {\it ibid}. 56, 3220 (1987); M. Ogata and Y. Wada, Jour. 
of the Phys. Soc. of Japan, 54,3425 (1985). 
\bibitem{neto} A. H. Castro Neto and A. O. Caldeira, Phys. Rev. B46, 8858,
(1992); A. H. Castro Neto and A. O. Caldeira, Phys. Rev. E48, 4037 (1993);
A. H. Castro Neto and A. O. Caldeira, Phys. Rev. A42, 6884 (1990).
\bibitem{b1} J. Schwinger, { J. of Math. Phys.} 2, 407 (1961).
\bibitem{maha} K. T. Mahanthappa, Phys. Rev. 126, 329 (1962); P. M.
Bakshi and K.T. Mahanthappa, J. Math. Phys. 4 1,12 (1963).
\bibitem{b2} L. V. Keldysh, {Sov. Phys. JETP} 20, 1018 (1965).
\bibitem{b3} V. Korenman, {Ann. Phys.} 39, 72 (1966).
\bibitem{b4} G. Z. Zhou, Z.B. Su, B. L. Hao and L. Yu., {Phys. Rep.} 118, 1 (1985).
\bibitem{b5} J. Rammer and H. Smith, Revs. of Mod. Phys. 58, 323, (1986).
\bibitem{b51} E. M. Lifshitz and L. P. Pitaevskii, ``Physical Kinetics'' 
(Pergamon, N.Y.) (1981); G. D. Mahan, ``Many Particle Physics''(2nd Edition) 
(Plenum, N.Y.) (1990); H. Kleinert, ``Path
Integrals in Quantum Mechanics, Statistics and Polymer Physics'' (2nd Edition),
World Scientific, Singapore, 1996);
R. Mills, ``Propagators for Many Particle 
Systems''  (Gordon and Breach, N. Y. 1969).
\bibitem{b6}  L. P. Kadanoff and G. Baym, ``Quantum 
Statistical Mechanics'' (Benjamin, N.Y. 1962).
\bibitem{b7} E. Calzetta and B. L. Hu, {Phys. Rev. D}35, 495 (1987); 
{Phys. Rev. D}37, 2878 (1988); {Phys. Rev. D}40, 656 (1989) and references therein.
\bibitem{c7} R. Rajaraman, `` Solitons and Instantons An Introduction to 
Solitons and Instantons in Quantum Field Theory'' (North-Holland Publishing Co.) 1982.
\bibitem{d2} J. Rubinstein, {J. Math. Phys.} 11 258 (1970).
\bibitem{c1} J. L. Gervais and B. Sakita, {Phys. Rev. D}11 2943 (1975).
\bibitem{c2} J. L. Gervais, A. Jevicki and B. Sakita, {Phys. Rev. D}12 1038 (1975).
\bibitem{c3} N. H. Christ and T. D. Lee, {Phys. Rev. D}12 1606 (1975).
\bibitem{c4} E. Tomboulis, {Phys. Rev. D}12 1678 (1975).
\bibitem{c5} J. L. Gervais and A. Jevicki, {Nucl. Phys.}B 93, 113 (1976).
\bibitem{c51} H. deVega, {Nucl. Phys. B} 115 411 (1976).
\bibitem{c6} T. D. Holstein and L. Turkevich, {Phys. Rev. B}38 1901 (1988);
{\it ibid} 1923; T. Holstein, Mol. Crys.Liq. Crys. 77,235 (1981).
\bibitem{c22} D. Jasnow and J. Rudnick, Phys. Rev. Lett. 41, 698 (1978); 
J. Rudnick and D. Jasnow, Phys. Rev. B 24, 2760 (1981). 
\bibitem{jackiw} R. Jackiw and J. R. Schrieffer, Nucl. Phys. B190, 253 (1981);
R. Jackiw and C. Rebbi, Phys. Rev. D13, 3398 (1976).
\bibitem{wil} J. Goldstone and F. Wilczek, Phys. Rev. Lett. 47, 986, (1981).
\bibitem{d5} R. Feynman and F. Vernon,  Ann. of Phys.(N.Y.) 24 118 (1963).
\bibitem{caldeira} A. O. Caldeira and A. J. Leggett, Physica A 121, 587
(1983). 
\bibitem{d6} A Schmid, {Jour. of Low Temp. Phys.} 49 609 (1982).
\bibitem{d7} H. Grabert, P. Schramm and G. -L. Ingold, Phys. Rep. 168, 115 (1988).
\bibitem{d8} U. Weiss, ``Quantum Dissipative Systems'', (World Scientific, 
Singapore, 1993), and references therein. 
\bibitem{negele} J. W. Negele and H. Orland, ``Quantum Many-Particle Systems''
(Addison-Wesley) 1988. 
\bibitem{d1} P. Morse and H. Feshbach, ``Methods of Mathematical Physics'', 
(McGraw-Hill), p.1650 (1953).
\end{thebibliography}
\end{document}